\documentclass[twocolumn]{autart}    

\usepackage[english]{babel}
\usepackage{amsmath,amssymb,latexsym,amsfonts,mathptmx}
\usepackage{xcolor,graphicx}

\usepackage{enumitem}
\usepackage{epsfig}
\usepackage{subfigure}
\usepackage{natbib}
\newcommand{\nin}{\not\in}

\newcommand{\tmtextit}[1]{{\itshape{#1}}}

\newcommand{\closure}{_{\text{cl}}}

\newcommand{\mylabel}[2]{#2\def\@currentlabel{#2}\label{#1}}
\newtheorem{theorem}{Theorem}[section]

\newtheorem{assumption}[theorem]{Assumption}

\newtheorem{lemma}[theorem]{Lemma}

\newtheorem{remark}[theorem]{Remark}

\newtheorem{proposition}[theorem]{Proposition}

\newcommand{\oprocendsymbol}{\hbox{$\bullet$}}
\newcommand{\oprocend}{\relax\ifmmode\else\unskip\hfill\fi\oprocendsymbol}

\newcommand{\real}{{\mathbb{R}}}
\newcommand{\naturals}{{\mathbb{N}}}

\renewcommand{\theenumi}{(\roman{enumi})}

\newcommand{\II}[1]{\mathcal{I}^{#1}}
\newcommand{\zeros}{\bold{0}}
\newcommand{\sat}{\text{sat}}
\newcommand{\ones}{\bold{1}}

\newcommand{\setdef}[2]{\left\{ #1 \; \big| \; #2\right\}}

\newcommand{\longthmtitle}[1]{\tmtextit{(#1).}}

\newcommand{\range}{\operatorname{range}}

\usepackage[normalem]{ulem}

\newcommand{\part}{\operatorname{par}}

\allowdisplaybreaks
\begin{document}

\begin{frontmatter}


\title{ Model predictive control for transient frequency
 \\ regulation
  of power networks}

%
%
%
%
%

\thanks[footnoteinfo]{During the preparation of this work, Y. Zhang
  was affiliated with the Department of Mechanical and Aerospace
  Engineering, UC San Diego.  This work was supported by NSF award
  CNS-1446891 and AFOSR Award FA9550-15-1-0108.  A preliminary version
  appeared as~\citep{YZ-JC:18-cdc2} at the IEEE Conference on Decision
  and Control.}

\author[YZ]{Yifu Zhang}\ead{yifu.zhang19@gmail.com}\hspace{2cm}
\author[JC]{Jorge Cort{\'e}s}\ead{cortes@ucsd.edu}

\address[YZ]{The MathWorks Inc., Natick, MA, 01760, USA}
\address[JC]{Department of Mechanical and Aerospace Engineering,
  University of California, San Diego, La Jolla, CA, 92093, USA}

\begin{keyword} 
  Power network stability, transient frequency, distributed control,
  model predictive control, convexification.
\end{keyword}

\begin{abstract}
  This paper introduces a control strategy to simultaneously achieve
  asymptotic stabilization and transient frequency regulation of power
  networks. The control command is generated by iteratively solving an
  open-loop control cost minimization problem with stability and
  transient frequency constraints. To deal with the non-convexity of
  the stability constraint, we propose a convexification strategy that
  uses a reference trajectory based on the system's current state.  We
  also detail how to employ network partitions to implement the
  proposed control strategy in a distributed way, where each region
  only requires system information from neighboring regions to execute
  its controller.
\end{abstract}

\end{frontmatter}

\section{Introduction}\label{section:intro}
To maintain system security and integrity~\citep{PK-JP:04}, power
networks are required to operate around their nominal frequencies in
the presence of disturbances, and recover synchronization as
disturbances disappear.  However, such a transient frequency
requirement faces fundamental challenges due to the deeper frequency
nadir caused by higher penetration of renewable generators with lower
inertia~\citep{FM-FD-GH-DJH-GV:2018,JF-HL-YT-FB:18}. This motivates our focus here on
developing methods to actively attenuate transient frequency
deviations while preserving network synchronization.
        
\textit{Literature review:} Work in~\citep{HDC:11,FD-MC-FB:13}
investigates power network synchronization conditions and their
relations to system dynamics and initial conditions. However, such
ideal conditions face challenges in practical scenarios with desired
safe limits that transient frequencies may violate. On the other hand,
various control schemes have been proposed to enhance transient
frequency behavior, including power dispatch~\citep{AA-EBM:06}, power
system stabilizer~\citep{PK:94}, feedback linearization
excitation~\citep{MAM-HRP-MA-MJH:14}, and virtual inertial
placement~\citep{TSB-TL-DJH:15}.  Nonetheless, these strategies do not
provide guarantees that the transient frequency will only evolve
within safe limits. To address this point, our previous
work~\citep{YZ-JC:19-auto} has combined Lyapunov stability and
invariance analysis to propose a distributed controller simultaneously
guaranteeing synchronization and transient frequency safety; however,
the proposed controller does not actively forecast the disturbance
evolution and its impact on transient frequency. As a result, it might
result in significant control efforts that could otherwise have been
avoided if the control action had been exerted earlier, something we
address here through a model predictive control (MPC) architecture.  A
related body of
work~\citep{ANV-IAH-JBR-SJW:08,DQM-JBR-CVR-POMS:00,DJ-BK:02} looks at
reducing control effort while respecting performance requirements, and
investigates distributed MPC for networked systems. However, the
proposed distributed implementations may jeopardize network stability.
Particularly,~\cite{DJ-BK:02} treats each subsystem as an independent
system by considering the effect of other subsystems as bounded
uncertainty, which complicates obtaining stability guarantees for the
whole system. In fact,~\cite{ANV-IAH-JBR-SJW:08} show that, if each
subsystem has no knowledge of other subsystems' cost
functions~\citep{EC-DJ-BHK-ST:02}, this leads to a noncooperative
game, and the control input trajectory may even diverge.  In addition,
some MPC approaches~\citep{ANV-IAH-JBR-SJW:08,MHN-etal:14} restrict
the predicted horizon to a single step in order to obtain distributed
strategies, since otherwise the control signal may require global
state or global system parameter information.
%

\textit{Statement of contribution:} This paper develops a distributed
receding-horizon control strategy that is able to simultaneously
maintain local asymptotic stability of the system and regulate
transient frequency. Specifically, for any given bus of interest, a
safe frequency region is both invariant and attractive under the
proposed design. For each state, we first formulate a non-convex
finite-horizon open-loop optimal control problem whose solution is the
control trajectory minimizing the overall cost under stability and
transient frequency constraints. We then propose a reference
trajectory technique for convexification.  The centralized closed-loop
control signal for each state is defined as the first-step solution of
the optimal control problem. To enable distributed control, we
partition the network into different regions and apply the centralized
control for each region, while taking into account the dynamics of
transmission lines connecting different regions. The resulting control
signal for each bus only relies on system information of the region to
which the bus belongs to and its neighboring regions.

%

\section{Problem statement}\label{section:ps}
In this section we introduce the model for the power network dynamics
and state the control goals\footnote{We use the following notation.
  $\naturals$, $\real$, $\real_{>}$, and $\real_{\geqslant}$ denote
  the set of natural, real, positive, and nonnegative real numbers,
  resp.  Variables are assumed to belong to Euclidean space if not
  specified otherwise. Let $\ones_n$ and $\zeros_n$ be the vector of
  all ones and zeros, resp. Denote $\partial\mathcal{Q}$ as the
  boundary of a set $\mathcal{Q}$.  We let $\lceil \cdot \rceil$
  denote the ceiling operator and $\|\cdot\|$ denote the 2-norm on
  $\real^{n}$. For any $c,d\in\naturals$, let $[c,d]_{\naturals}=
  \left\{ x\in\naturals \big| c\leqslant x\leqslant d \right\}$. For
  $\mu\in\{0,1\}$ and $a^{\min}<a^{\max}$, the saturation function is
  $\sat (a;\mu,a^{\min},a^{\max}) = a^{\min}$ if $\mu=0$ and
  $a\leqslant a^{\min}$, $\sat (a;\mu,a^{\min},a^{\max}) = a^{\max}$
  if $\mu=0$ and $a\geqslant a^{\max}$, and $\sat
  (a;\mu,a^{\min},a^{\max}) = a$ otherwise.  For $b\in\real^{n}$,
  $b_{i}$ denotes its $i$th entry and for $A\in\mathbb{R}^{m\times
    n}$, $[A]_i$ and $[A]_{i,j}$ denote its $i$th row and $(i,j)$th
  element. We denote by $A^{\dagger}$ and $\range(A)$ its unique
  Moore-Penrose pseudoinverse and column space, resp.  }.  Consider a
power network described by a connected undirected graph,
cf.~\citep{FB-JC-SM:08cor}, $\mathcal{G}=(\mathcal{I},\mathcal{E})$,
where $\mathcal{I}=\{1,2,\cdots,n\}$ is the collection of buses and
$\mathcal{E}=\{e_{1},e_{2},\cdots,e_{m}\}\subseteq\mathcal{I}\times\mathcal{I}$
is the collection of transmission lines.  For each node
$i\in\mathcal{I}$, let $M_{i}\in\real_{\geqslant}$,
$E_{i}\in\real_{>}$, $\omega_{i}\in\real$ and $p_{i}\in\real$ denote
its inertia, damping coefficient, shifted voltage frequency relative
to the nominal frequency, and active power injection, resp. Note that
we explicitly allow some buses to have zero inertia, and we assume
that at least one bus possesses strictly positive inertia.  For
compactness, define $M \triangleq
\text{diag}(M_{1},M_{2},\cdots,M_{n})\in\real^{n\times n}$, $E
\triangleq \text{diag}(E_{1},E_{2},\cdots,E_{n})\in\real^{n\times n}$,
$\omega\triangleq(\omega_{1},\omega_{2},\cdots,
\omega_{n})^{T}\in\real^{n}$ and $p\triangleq (p_{1},p_{2},\cdots,
p_{n})^{T}\in\real^{n}$. For each edge $e_{k} \in \mathcal{E}$ with
vertices $i$, $j$, an orientation consists of choosing one node, say
$i$, to be the positive end of $e_{k}$ and the other vertex, $j$, to
be the negative end.  Let $D=(d_{k i}) \in \mathbb{R}^{m \times n}$ be
the incidence matrix corresponding to the chosen orientation (i.e., $
d_{k i} = 1$ if $i$ is the positive end of $e_{k}$, $ d_{k i} = -1$ if
$i$ is the negative end of $e_{k}$, and $ d_{k i} = 0$
otherwise). {Two nodes $i$ and $j$ are neighbors if there is an edge
  connecting them, and we} let $\lambda_{ij}$ denote the voltage angle
difference between $i$ and~$j$.
Let $\lambda\in\real^{m}$ denote the collection of $\lambda_{ij}$ and
$Y_{b}\in\real^{m\times m}$ be the diagonal matrix whose $k$th entry
represents the susceptance of the transmission line $e_{k}$ connecting
bus $i$ and $j$, i.e., $[Y_{b}]_{k,k}=b_{ij},$ for $ k=1,2,\cdots, m$.
We partition buses into $\II{u}$ and $\mathcal{I}\backslash\II{u}$,
depending on whether an additional control input is available to
regulate transient frequency behavior.  The swing
equations~\citep{JM-JWB-JRB:08} describe the evolution of voltage angle
difference and frequencies as
 \begin{subequations}\label{eqn:compact-form}
  \begin{align}
    \dot \lambda(t)&=D\omega(t),
    \\
    M\dot\omega(t)&=-E\omega(t)-D^{T}Y_{
      b}\sin\lambda(t)+p(t)+u(t),\label{eqn:compact-form-2}
    \\
    & \hspace{-0.9cm}u(t)\in\mathbb{U}\triangleq\left\{ u\in\real^{n}
      \big| \ \forall w\in[1,n]_{\naturals},\ [u]_{w}=\left\{
        \hspace{-.2cm}\begin{array}{ccc} & u_{w} & \text{if
            $w\in\II{u}$} \notag
          \\
          & 0 & \text{otherwise}
    \end{array} \hspace{-0.2cm}\right. \right\},
  \end{align}
\end{subequations}
%
where $\sin\lambda(t)\in\real^{m}$ is taken component-wise.
For convenience, we use $ x\triangleq (\lambda,\omega)\in\real^{m+n}$
to denote the collection of all states. Note
that~\eqref{eqn:compact-form} is in fact a set of
differential-algebraic equations if at least one node has zero
inertia. In addition, although in the model we generally consider a
time-varying power injection $p$, some results developed later depend
on a stricter assumption stated as follows.

%

\begin{assumption}\longthmtitle{Time-invariant power
    injection}\label{assumption:TL-power}
  The power injection is constant, i.e., $p(t)=p^{*}\in\real$ for
  all $t\geqslant 0$.
\end{assumption}
%
%
Under this assumption, let
$\omega^{\infty} \triangleq \frac{\sum_{i=1}^{n}
  p_{i}^{*}}{\sum_{i=1}^{n}E_{i}}$ and
$\tilde p=p^{*}-\omega_{\infty}E\ones_{n}$.  Consider
$L\triangleq D^{T}Y_{b}D$ the Laplacian matrix of the network graph
and define
$\| z \|_{\mathcal{E},\infty} \triangleq \max_{(i,j)\in\mathcal{E}}
|z_{i}-z_{j}|$ for vector $z\in\real^{n}$.  Then, one can
show~\citep[Lemma 2 and inequality (S17)]{FD-MC-FB:13} that, for the
system~\eqref{eqn:compact-form} with $u \equiv 0_{n}$, if
\begin{align}\label{ineq:sufficient-eq}
  \|L^{\dagger}\tilde p\|_{\mathcal{E},\infty}<1,
\end{align}
then there exists an equilibrium point
$(\lambda^{\infty},\omega^{\infty}\ones_{n})\in\real^{m+n}$ that is
locally asymptotically stable. Specifically,
$\lambda^{\infty}\in\Upsilon$ and is unique in its closure
$\Upsilon\closure$, where
$\Upsilon\triangleq\setdef{\lambda}{|\lambda_{i}|<\pi/2,\ \forall
  i\in[1,m]_{\naturals}}$.
The term
$\|L^{\dagger}\tilde p\|_{\mathcal{E},\infty}$ represents the
maximum steady-state voltage angle difference between adjacent nodes
for the linearized dynamics of~\eqref{eqn:compact-form} by replacing $\sin\lambda$ by $\lambda$.

         
We aim to design state-feedback controllers $u_{i}$ for each bus
$i\in\II{u}$ that stabilize the system, cooperatively ensure that the
frequencies of a targeted set of buses stay within safe bounds, and
force them to enter the safe bounds if they are initially
outside. We next list these requirements formally.

\emph{Safe frequency invariance requirement:} Given
$\II{\omega}\subseteq\II{u}$, for each $i\in\II{\omega}$, let
$\underline\omega_{i}, \bar\omega_{i}\in\real$ with
$\underline\omega_{i}<\bar\omega_{i}$ be lower and upper safe
frequency bounds. We require that the interval
$[\underline\omega_{i},\bar\omega_{i}]$ is invariant and attractive:
if $\omega_{i}(0)\in [\underline\omega_{i},\bar\omega_{i}]$, then
$\omega_{i}(t)\in [\underline\omega_{i},\bar\omega_{i}]$ for every
$t>0$ and, if $\omega_{i}(0) \nin
[\underline\omega_{i},\bar\omega_{i}]$, then $\omega_{i}$ enters the
interval in finite time, never to leave it afterwards.

\emph{Asymptotic stability requirement:} We require that the
controller only shapes transients so that the
$(\lambda^{\infty},\omega^{\infty}\ones_{n})$ remains locally
asymptotically stable for the closed-loop system.

%
\emph{Coordination requirement:} Each controller $u_{i}$,
$i\in\II{u}$, should cooperate with others to lower the overall
control effort, as measured by some given cost function.
        


Our design strategy is to first set up an open-loop optimization
problem with control cost as objective function, and with frequency
and stability requirements as constraints.  Then, we design a
centralized controller by solving this optimization problem in a
receding horizon fashion. Finally, the distributed controller comes
from partitioning the network into several regions, and treating each
region as an independent network.
%

\section{Open-loop optimal control}
We start by formulating an optimization problem whose goal is to
minimize a cost function measuring control input effort subject to the
system dynamics, safe frequency invariance, and asymptotic stability
constraints. As this problem turns out to be non-convex and
non-smooth, we propose a convexification strategy by generating a set
of linear constraints. Later, we build on this to design centralized
and distributed controllers.
        
\subsection{Open-loop finite-horizon optimal
  control}\label{subsection:option-loop}
We introduce a robust asymptotic stability condition with respect to
the open-loop equilibrium point and estimate the region of
attraction. {Let $\mathfrak{G}\in\mathcal{I}$ denote the collection
  of node indexes with strictly positive inertia, and
  $\omega_{g}\in\real^{|\mathfrak{G}|}$ be the corresponding
  collection of frequencies of these nodes.}  Consider the energy
function~\citep{YZ-JC:19-auto,TLV-HDN-AM-JS-KT:18,NM-CP:17}
\begin{align*}
  V(\lambda,\omega_{g})\triangleq\frac{1}{2}\sum_{i\in\mathfrak{G}}
  M_{i}(\omega_{i}-\omega^{\infty})^{2}+\sum_{j=1}^{m}[Y_{b}]_{j,j}a(\lambda_{j},
  \lambda_{j}^{\infty}),
\end{align*}
where $a(\lambda_{j},\lambda_{j}^{\infty}) \triangleq
\cos\lambda_{j}^{\infty}-\cos\lambda_{j} -
\lambda_{j}\sin\lambda_{j}^{\infty} +
\lambda_{j}^{\infty}\sin\lambda_{j}^{\infty}$. Furthermore, let $\bar
r\triangleq\min_{\tilde \lambda\in\partial\Upsilon_{cl}}
V(\tilde\lambda,\omega^{\infty}\ones_{|\mathfrak{G}|})$. Roughly
speaking, the first and second terms in $V$ represent the stored
kinetic energy and elastic potential energy, respectively.  The
following result is a generalization of~\citep[Lemma
4.1]{YZ-JC:19-auto}.

\begin{lemma}\longthmtitle{Robust asymptotic stability
    condition} \label{prop:robust-as}
  For system~\eqref{eqn:compact-form}, suppose that the solution
  exists and is unique.  For every $i\in\II{u}$, let
  $\bar\omega_{i}^{\text{thr}}>0$ and $
  \underline\omega_{i}^{\text{thr}}<0$ be threshold values satisfying
  $\underline\omega_{i}^{\text{thr}}<\omega^{\infty}<\bar\omega_{i}^{\text{thr}}$. If
  for every $t\in\real_{\geqslant}$,
  \begin{subequations}\label{ineq:robust-stabilize-constraints}
    \begin{align}
      \omega_{i}(t)u_{i}(x(t),p(t))&\leqslant 0, \ \text{if
      }\omega_{i}(t) \nin
      (\underline\omega_{i}^{\text{thr}},\bar\omega_{i}^{\text{thr}}), 
      \label{ineq:robust-stabilize-constraints-2a}
      \\
      u_{i}(x(t),p(t))&=0, \ \text{if }\omega_{i}(t)\in
      (\underline\omega_{i}^{\text{thr}},\bar\omega_{i}^{\text{thr}}), 
      \label{ineq:robust-stabilize-constraints-2b}
    \end{align}
  \end{subequations}
  then under Assumption~\ref{assumption:TL-power} and
  condition~\eqref{ineq:sufficient-eq},
  $(\lambda^{\infty},\omega^{\infty}\ones_{n})$ is locally
  asymptotically stable. Furthermore, define
  \begin{align}\label{set:region}
    \Phi(r)\triangleq\setdef{(\lambda,\omega_{g})}{\lambda\in
      \Upsilon\closure,\ V(\lambda,\omega_{g})\leqslant r}.
  \end{align}
  Then for every $(\lambda(0),\omega_{g}(0))\in\Phi(r)$ with $0<r<\bar
  r$, it holds that $(\lambda(t),\omega_{g}(t))\in\Phi(r)$ for every
  $t\geqslant 0$ and $(\lambda(t),\omega(t))
  \rightarrow(\lambda^{\infty},\omega^{\infty}\ones_{n})$.
\end{lemma}
\begin{pf}
  We prove that if $p(t)\equiv p^{*}$,
  then~\eqref{ineq:robust-stabilize-constraints} implies
  \begin{subequations}\label{ineq:stabilize-constraints-2}
    \begin{align}
      (\omega_{i}(t)-\omega^{\infty})u_{i}(x(t),p^{*})&\leqslant 0, \
      \text{if }
      \omega_{i}(t)\neq\omega^{\infty},\label{ineq:stabilize-constraints-2a}
      \\
      u_{i}(x(t),p^{*})&=0, \ \text{if
      }\omega_{i}(t)=\omega^{\infty}.\label{ineq:stabilize-constraints-2b}
    \end{align}
  \end{subequations}
  If $\omega_{i}(t) >\bar\omega_{i}^{\text{thr}}$,
  then~\eqref{ineq:robust-stabilize-constraints} is equivalent to
  asking $u_{i}(x(t),p^{*})\geqslant 0$, which
  guarantees~\eqref{ineq:stabilize-constraints-2} by noticing
  $\omega_{i}(t)-\omega^{\infty}>\bar\omega_{i}^{\text{thr}}-\omega^{\infty}>0$. A
  similar argument works when $\omega_{i}(t)
  <\underline\omega_{i}^{\text{thr}}$. Finally, if $\omega_{i}(t)\in
  (\underline\omega_{i}^{\text{thr}},\bar\omega_{i}^{\text{thr}})$,
  then~\eqref{ineq:robust-stabilize-constraints} requires
  $u_{i}(x(t),p^{*})= 0$,
  ensuring~\eqref{ineq:stabilize-constraints-2}. {
    From~\citep[Theorem~1]{NM-CP:17}, one has
    \begin{align}\label{ineq:dotV}
      \dot
      V(\lambda(t),\omega_{g}(t))&=-\sum_{i\in\mathfrak{G}}E_{i}(\omega_{i}(t)-\omega^{\infty})^{2}-\hspace{-.3cm}\sum_{i\in\II{u}\slash\mathfrak{G}}E_{i}(\omega_{i}(t)-\omega^{\infty})^{2}\notag
      \\&-\sum_{i\in\II{u}}(\omega_{i}(t)-\omega_{\infty})u_{i}(x(t),p^{*})\leqslant0,
    \end{align}  
    where $\omega_{i}(t)$ with $i\in\II{u}\slash\mathfrak{G}$ is a
    function of $(\lambda(t),\omega_{g}(t))$. Specifically,
    by~\eqref{eqn:compact-form-2}, one has
    $E_{i}\omega_{i}(t)=-[D^{T}Y_{b}]_{i}\sin\lambda(t)+p_{i}^{*}+u_{i}(t)$.
    Furthermore,
    one has that~i) $\Phi(r)$ is compact and non-empty, ii)
    $V(\lambda,\omega_{g})\geqslant 0$ for every
    $(\lambda,\omega_{g})\in\Phi(r)$, and the equality holds only when
    $(\lambda,\omega_{g})=(\lambda^{\infty},\omega^{\infty}\ones_{|\mathfrak{G}|})$. These
    two properties, together with~\eqref{ineq:dotV}, imply the
    convergence of $(\lambda,\omega)$ to
    $(\lambda^{\infty},\omega^{\infty}\ones_n)$ by the LaSalle
    Invariance Principle~\citep[Theorem~4.4]{HKK:02}.  } \qed
\end{pf}

Notice that the dependence of the robust asymptotic stability
condition~\eqref{ineq:robust-stabilize-constraints} on the equilibrium
point $(\lambda^{\infty},\omega^{\infty}\ones_{n})$ is limited to an
approximate knowledge of $\omega^{\infty}$.  This reflects a practical
consideration under which the controller should still ensure
asymptotic stability: although ideally $\omega^{\infty}$ is 0 when
load and supply are balanced (i.e., $\sum_{i=1}^{n}p_{i}^{*}=0$), due
to imperfect estimation on the load side and transmission losses,
$\omega^{\infty}$ tends to slightly deviate from $0$.

With the stability condition being set, we now are ready to formally
introduce the finite-horizon optimal control problem.
As the power injection $p$ may not be precisely predicted a priori,
instead, for every $t\in\real_{\geqslant}$, we consider a piece-wise
continuous signal $p^{fcst}_{t}:[t,t+\tilde t]\rightarrow\real^{n}$
forecasting its value for the first $\tilde t$ seconds starting
from~$t$. When convenient, we invoke the following assumption in our
technical analysis.

\begin{assumption}\longthmtitle{Forecast reveals  true
    value at current time}\label{assumption:forecast-injection}
  For any $t\in\real_{\geqslant }$, $p^{fcst}_{t}(t)=p(t)$.
\end{assumption}


The open-loop finite-horizon optimal control problem is defined
in~\eqref{opti:nonlinear-continuous},
\begin{figure*}[htb]
  \begin{subequations}\label{opti:nonlinear-continuous}
    \begin{alignat}{3}
      (Q_{cont}) \hspace{10mm} & \min_{\lambda,\omega,u,\beta,\gamma}
      &\quad & \sum_{i\in\II{u}}\int_{\tau_{0}}^{\tau_{0}+\tilde
        t}c_{i}u^{2}_{i}(\tau)+d_{i}\beta_{i}^{2}(\tau)\text{d}\tau+\sum_{i\in\II{\omega}}\int_{\tau_{0}}^{\tau_{0}+\tilde
        t}e_{i}\gamma_{i}^{2}(\tau)\text{d}\tau && \notag
      \\
      &\text{s.t.}&\quad &\dot \lambda(\tau)=D\omega(\tau),
      && \label{opti:nonlinear-a}
      \\
      &&& M\dot\omega(\tau)=-E\omega(\tau)-D^{T}Y_{
        b}\lambda(\tau)+p^{fcst}_{t}(\tau)+u(\tau),
      && \label{opti:nonlinear-b}
      \\
      &&&\lambda(\tau_{0})=\sin \lambda_{ 0} ,\;
      \omega(\tau_{0})=\omega_{0}, && \label{opti:nonlinear-c}
      \\
      &&&u(\tau)\in\mathbb{U},\quad && \forall
      \tau\in[\tau_{0},\tau_{0}+\tilde t],\label{opti:nonlinear-d}
      \\
      &&& u_{i}^{\min}-\xi_{i}\beta_{i}(\tau) \leqslant
      u_{i}(\tau)\leqslant u_{i}^{\max}+\xi_{i}\beta_{i}(\tau), &&
      \forall i\in\II{u},\;\forall \tau\in[\tau_{0},\tau_{0}+\tilde
      t],\label{opti:nonlinear-e}
      \\
      &&& \beta_{i}(\tau)\geqslant 0, && \forall i\in\II{u},\ \forall
      \tau\in[\tau_{0},\tau_{0}+\tilde t],\label{opti:nonlinear-f}
      \\
      &&&
      \underline\omega_{i}-\kappa_{i}(\omega_{0},\xi_{i})(\gamma_{i}(\tau)-\delta)\leqslant
      \omega_{i}(\tau)\leqslant
      \bar\omega_{i}+\kappa_{i}(\omega_{0},\xi_{i})(\gamma_{i}(\tau)-\delta), \quad
      && \forall i\in\II{\omega},\;\forall
      \tau\in[\tau_{0},\tau_{0}+\tilde
      t],\label{ineq:hard-frequency-constraint}
      \\
      &&& \gamma_{i}(\tau)\geqslant 0, && \forall i\in\II{\omega}, \
      \forall \tau\in[\tau_{0},\tau_{0}+\tilde
      t],\label{ineq:hard-frequency-constraint-2}
      \\
      &&&
      (\omega,u)\in\Phi_{cont}, && \label{set:opti-nonlinear-continuous}
    \end{alignat}
  \end{subequations}
  \hrulefill
\vspace*{-2ex}
\end{figure*}
where constraints~\eqref{opti:nonlinear-a}-\eqref{opti:nonlinear-c}
represent system dynamics and initial state. Notice that we linearize
the dynamics in~\eqref{opti:nonlinear-b}, which contributes to the
convexification of the open-loop optimization with a slight loss of
optimality (in Section~\ref{sec:closed-loop-analysis}, we show that
employing this linearization for controller design does not jeopardize
the asymptotic stability or safe frequency invariance requirements in
the closed-loop system); constraint~\eqref{opti:nonlinear-d} reflects
the availability of control signal at each node;
constraints~\eqref{opti:nonlinear-e} and~\eqref{opti:nonlinear-f}
delimit the control magnitude bounds, in which $\xi\in\{0,1\}$
indicates the magnitude constraint type, i.e., if $\xi_{i}=1$ for
$i\in\II{u}$, then the constraint is soft as $u_{i}(\tau)$ could
exceed $u_{i}^{\max}\in\real$ or $u_{i}^{\min}\in\real$, but penalized
by $\beta_{i}(\tau)$ in the objective function, and if $\xi_{i}=0$
then it is a hard constraint;
constraints~\eqref{ineq:hard-frequency-constraint}
and~\eqref{ineq:hard-frequency-constraint-2} refer to the safe
frequency invariance requirement, in which
\begin{align}
  \kappa_{i}(\omega_{0},\xi_{i})=
  \begin{cases}
    0 & \hspace{0.5cm}\text{if $\omega_{i,0}\in[\underline
      \omega_{i},\bar\omega_{i}]$ and $\xi_{i}=1$,}
    \\
    1 & \hspace{2.9cm}\text{otherwise.}
  \end{cases}
\end{align}
Intuitively, these two constraints require that $\omega_{i}$ stays in
$[\underline \omega_{i},\bar\omega_{i}]$ provided that it is initially
inside and the magnitude constraint on the controller is soft, and
penalize through $\gamma_{i}$ if not. The parameter$\delta_i$ with
$0<\delta_{i}<\bar\omega_{i}-\underline\omega_{i}$ is tunable, forcing
$\omega_{i}(\tau)$ approach the interval $[\underline
\omega_{i}+\delta_{i},\bar\omega_{i}-\delta]$, and hence enter
$[\underline \omega_{i},\bar\omega_{i}]$ in finite time;
constraint~\eqref{set:opti-nonlinear-continuous} is the asymptotic
stability condition established in Lemma~\ref{prop:robust-as}, where
\begin{align*}
  \Phi_{cont}\triangleq\left\{ (\omega,u) \; \big| \;
    \eqref{ineq:robust-stabilize-constraints}\text{ holds }\forall
    t\in[\tau_{0}.\tau_{0}+\tilde t],\;\forall i\in \II{u}\right\}.
\end{align*}
Finally, $c_{i},d_{i},e_{i}\in\real_{>}$ 
refer to the weight coefficient on control effort, control magnitude
penalty, and frequency invariance penalty, resp.

We refer to~\eqref{opti:nonlinear-continuous} as $Q_{cont}(\mathcal{G}
,\II{u},\II{\omega},p_{t}^{fcst},\lambda_{0},\omega_{0},\tau_{0})$ to
emphasize its dependence on the graph topology, controlled node
indexes, transient-frequency-constrained node indexes, forecasted
power injection, initial state, and initial time. If the context is
clear, we use~$Q_{cont}$. We use the same notational logic for other
optimization problems in the rest of the paper.

In practice, a convenient way to approximate the functional solution
for $Q_{cont}$ is by discretization. Specially, here we discretize the
system periodically with time length $T\in\real_{>}$, and denote
$N\triangleq \lceil\tilde t/T\rceil$ as the total number of steps.
For every $k\in[0,N]_{\naturals}$, denote $\hat
\lambda(k),\hat\omega(k)$, $\hat u(k)$, $\hat p^{fcst}(k)$ as the
approximation of $\lambda(\tau_{0}+kT),\omega(\tau_{0}+kT)$,
$u(\tau_{0}+kT)$ and $p^{fcst}_{t}(\tau_{0}+kT)$, resp., and let
\begin{subequations}\label{sube:eqn:traj}
  \begin{align}
    \hat \Lambda&\triangleq[\hat \lambda(0),\hat \lambda(1),\cdots,
    \lambda(N)],
    \\
    \hat
    \Omega&\triangleq[\hat\omega(0),\hat\omega(1),\cdots,\hat\omega(N)],
    \\
    \hat P^{fcst}&\triangleq[\hat p^{fcst}(0),\hat
    p^{fcst}(1),\cdots,\hat p^{fcst}(N-1)],\label{eqn:p-G-graph}
    \\
    \hat U&\triangleq[\hat u(0),\hat u(1),\cdots,\hat u(N-1)],
    \\
    \hat B&\triangleq
    [\hat\beta(0),\hat\beta(1),\cdots,\hat\beta(N-1)],
    \\
    \hat \Gamma&\triangleq
    [\hat\gamma(0),\hat\gamma(1),\cdots,\hat\gamma(N)],
  \end{align}
\end{subequations}
be the collection of voltage angle difference, frequency, predicted power injection,
and control input discrete trajectories, resp. We formulate
the discrete version of $Q_{disc}$ in~\eqref{opti:nonlinear}, where
\begin{align}
  \Phi_{disc}\triangleq\Big\{ (\hat\Omega,\hat U)\; \big| \; \forall
  i\in\II{u},\ \forall k\in[0,N-1]_{\naturals}, \text{ it holds that}
  \notag\\
  & \hspace{-6.9cm}\hat\omega_{i}(k)\hat u_{i}(k)\leqslant 0, \
  \text{if }\hat\omega_{i}(k)\nin
  (\underline\omega_{i}^{\text{thr}},\bar\omega_{i}^{\text{thr}}),\notag
  \\
  &\hspace{ -5.2cm} \hat u_{i}(k)=0, \ \text{if }\hat\omega_{i}(k)\in
  (\underline\omega_{i}^{\text{thr}},\bar\omega_{i}^{\text{thr}})\Big\}.
\end{align}
Note that this set is nonlinear and non-smooth.

\begin{figure*}[htb]
  \begin{subequations}\label{opti:nonlinear}
    \begin{alignat}{3}
      (Q_{disc})  \hspace{10mm} & \min_{\hat \Lambda,\hat\Omega, \hat
        U,\hat B,\hat\Gamma} & \quad & g(\hat U,\hat
      B,\hat\Gamma)\triangleq
      \sum_{i\in\II{u}}\sum_{k=0}^{N-1}\left(c_{i}\hat
        u^{2}_{i}(k)+d_{i}\beta_{i}^{2}(k)\right)+
      \sum_{i\in\II{\omega}}\sum_{k=1}^{N}e_{i}\gamma_{i}^{2}(k) &&
      \notag
      \\
      &\text{s.t.}&\quad &\hat \lambda(k+1)=\hat
      \lambda(k)+TD\hat\omega(k), && \notag
      \\
      &&&{M(\hat\omega(k+1)-\hat\omega(k))\slash T}=-E\hat\omega(k)-D^{T}Y_{b}\hat
      \lambda(k)+\hat p^{fcst}(k)+\hat u(k), \quad && \forall k
      \in[0,N-1]_{\naturals},\label{opti:nonlinear-1}
      \\
      &&&\hat \lambda(0)=\sin \lambda_{ 0},\
      \hat\omega(0)=\omega_{0}, && \label{opti:nonlinear-2}
      \\
      &&&\hat u(k)\in\mathbb{U}, &&\forall k
      \in[0,N-1]_{\naturals},\label{opti:nonlinear-2b}
      \\
      &&& u_{i}^{\min}-\xi_{i}\beta_{i}(k) \leqslant \hat
      u_{i}(k)\leqslant u_{i}^{\max}+\xi_{i}\beta_{i}(k),
      && \forall i\in\II{u},\;\forall
      k\in[0,N-1]_{\naturals},\label{opti:nonlinear-4}
      \\
      &&& \beta_{i}(k)\geqslant 0,  && \forall i\in\II{u},
      \forall k\in[0,N-1]_{\naturals},\label{opti:nonlinear-5}
      \\
      &&&
      \underline\omega_{i}-\kappa_{i}(\omega_{0},\xi_{i})(\gamma_{i}(k)-\delta)\leqslant
      \hat\omega_{i}(k)\leqslant
      \bar\omega_{i}+\kappa_{i}(\omega_{0},\xi_{i})(\gamma_{i}(k)-\delta),
      &&  \forall i\in\II{\omega},\ \forall
      k\in[1,N]_{\naturals},\hspace{-0.2cm}\label{opti:nonlinear-3}
      \\
      &&& \gamma_{i}(k)\geqslant 0, && \forall
      i\in\II{\omega}, \forall
      k\in[1,N]_{\naturals},\label{opti:nonlinear-6}
      \\
      &&& (\hat\Omega,\hat U)\in\Phi_{disc}, &&\label{set:opti-nonlinear}
  \end{alignat}
\end{subequations}
\hrulefill
\vspace*{-2ex}
\end{figure*}

\subsection{Constraint convexification}
The major obstacle to solve $Q_{disc}$ is dealing with the
set~$\Phi_{disc}$ in constraint~\eqref{set:opti-nonlinear}\footnote{
  In fact, the non-smoothness of the set $\Phi_{disc}$ makes standard
  methods in nonlinear optimization (e.g., interior point method,
  sequential quadratic programming, trust region method) occasionally
  fail to return even a feasible solution (let along a local
  optimizer) since they require the existence of a gradient for
  every constraint~\citep{fmincon}.}.  To this end, we propose a
convexification method that seeks to identify a subset of
$\Phi_{disc}$ consisting of only linear constraints. This method
relies on the notion of \emph{reference trajectory}, which is a
trajectory $(\hat \Lambda,\hat\Omega,\hat U)$ of the system state and
input for which there exist $\hat B $ and $\hat\Gamma$ such
that~\eqref{opti:nonlinear} are satisfied.
The next result details this.

\begin{lemma}\longthmtitle{Convexification of non-convex
    constraints} \label{lemma:convexificaiton}
  For any reference trajectory $(\hat
  \Lambda^{\text{ref}},\hat\Omega^{\text{ref}},\hat U^{\text{ref}})$, let
  \begin{align}
    \Phi_{cvx}\triangleq\Big\{ (\hat\Omega,\hat U)\; \big| \; \forall
    i\in\II{u},\ \forall k\in[0,N-1]_{\naturals}, \text{ it holds
      that} \notag
    \\
    &\hspace{-7.2cm}\hat\omega_{i}(k)\geqslant
    \bar\omega_{i}^{\text{thr}},\ \hat u_{i}(k)\leqslant 0, \ \text{if
    }\hat\omega_{i}^{\text{ref}}(k)\geqslant
    \bar\omega_{i}^{\text{thr}};\notag
    \\
    &\hspace{-7.2cm}\hat\omega_{i}(k)\leqslant
    \underline\omega_{i}^{\text{thr}},\ \hat u_{i}(k)\geqslant 0, \
    \text{if }\hat\omega_{i}^{\text{ref}}(k)\leqslant
    \underline\omega_{i}^{\text{thr}};\notag
    \\
    &\hspace{ -6.5cm} \hat u_{i}(k)=0, \ \text{if
    }\underline\omega_{i}^{\text{thr}} <
    \hat\omega_{i}^{\text{ref}}(k)< \bar\omega_{i}^{\text{thr}}\Big\}.
  \end{align}
  Then, $\Phi_{cvx}$ is convex and satisfies $\emptyset \neq
  \Phi_{cvx}\subseteq\Phi_{disc}$.
\end{lemma} 
\begin{pf}
  The non-emptiness holds by simply noticing that
  $(\hat\Omega^{\text{ref}},\hat U^{\text{ref}})\in\Phi_{cvx}$. We
  show the inclusion by classifying each $k\in[0,N-1]_{\naturals}$
  into three types regarding the value of
  $\hat\omega_{i}^{\text{ref}}(k)$.  If
  $\hat\omega_{i}^{\text{ref}}(k)\geqslant
  \bar\omega_{i}^{\text{thr}}$, then at step $k$, only the first
  constraint in $\Phi_{cvx}$ is active, which satisfies the first
  constraint in $\Phi_{disc}$, as well as the second one trivially,
  since in this case $\hat \omega_{i}(k) \notin
  (\underline\omega_{i}^{\text{thr}},\bar\omega_{i}^{\text{thr}})$. Similar
  analysis holds if $\hat\omega_{i}^{\text{ref}}(k)\leqslant
  \underline\omega_{i}^{\text{thr}}$. Finally, if
  $\underline\omega_{i}^{\text{thr}} < \hat\omega_{i}^{\text{ref}}(k)<
  \bar\omega_{i}^{\text{thr}}$, then only the last constraint in
  $\Phi_{cvx}$ is active, which satisfies both two constraints in
  $\Phi_{disc}$. Finally, the convexity of $\Phi_{cvx}$ follows by
  noting that it corresponds to the intersection of finitely many
  linear constraints over all $i\in\II{u}$ and
  $k\in[0,N-1]_{\naturals}$.  To see this, notice that for each $i$
  and $k$, as the value of $\hat\omega_{i}^{\text{ref}}(k)$ is given a
  priori by the reference trajectory, one and only one of the three
  constraints in $\Phi_{cvx}$ is active, leading to linearity. \qed
\end{pf}
        
In light of Lemma~\ref{lemma:convexificaiton}, given a reference
trajectory, we solve a convexified version of $Q_{disc}$, replacing
$\Phi_{disc}$ by $\Phi_{cvx}$,
\begin{subequations}\label{opti:linear}
  \begin{alignat}{2}
    (Q_{cvx}) \hspace{10mm} & \min_{\hat F,\hat\Omega, \hat U} & \quad &
    g(\hat U,\hat B,\hat\Gamma)\notag
    \\
    &\text{s.t.}&\quad
    &~\eqref{opti:nonlinear-1}-\eqref{opti:nonlinear-6}\text{ hold},
    \\
    &&& (\hat\Omega,\hat U)\in\Phi_{cvx}.\label{set:opti-linear-phi}
  \end{alignat}
\end{subequations}
Since the convexification reduces the set $\Phi_{disc}$ to
$\Phi_{cvx}$, the optimal value of $Q_{disc}$ is less than or equal to
that of $Q_{cvx}$.  For consistency, if the reference trajectory is
the optimal solution of $Q_{disc}$, then both problems have the same
optimal value.

\subsection{Generation of reference trajectory}
Here we introduce a method to generate the reference trajectory
required by the convexification process of $\Phi_{disc}$. Our next
result shows that the discretization~\eqref{eqn:ref-controller} of the
continuous-time feedback controller designed
in~\citep[equation~(16)]{YZ-JC:19-auto} (which is able to guarantee
safe frequency invariant requirement for the continuous-time
system~\eqref{eqn:compact-form}) generates a valid reference
trajectory for the discretized system~\eqref{opti:nonlinear-1}.

\begin{proposition}\longthmtitle{Generation of reference
    trajectory}\label{prop:ref-generation}
  For every $i\in\II{u}$ and every $k\in[0,N-1]_{\naturals}$, suppose
  $\underline\omega_{i}<\underline\omega_{i}^{\text{thr}}<\omega^{\infty}
  < \bar\omega_{i}^{\text{thr}}<\bar\omega_{i}$, and $\bar\gamma_{i},\
  \underline\gamma_{i}\in\real_{>}$.  Let $\hat u^{\text{ref}}$ be
  defined as in~\eqref{eqn:ref-controller} and set $\hat
  U^{\text{ref}}\triangleq[\hat u^{\text{ref}}(0),\hat
  u^{\text{ref}}(1),\cdots,\hat u^{\text{ref}}(N-1)]$. Let 
  $(\hat\Lambda^{\text{ref}},\hat \Omega^{\text{ref}})$ be the sate
  trajectory uniquely determined by~\eqref{opti:nonlinear-1}
  and~\eqref{opti:nonlinear-2} using $\hat u^{\text{ref}}$ as input.
  Then there exists $\bar T\in\real_{>}$ such that for any
  $0<T\leqslant \bar T$, $(\hat
  \Lambda^{\text{ref}},\hat\Omega^{\text{ref}},\hat U^{\text{ref}})$
  is a reference trajectory.
\end{proposition}

\begin{figure*}[htb]
  \begin{align}\label{eqn:ref-controller}
    \hat u^{a}_{i}(k)&\triangleq\left\{ \begin{array}{ccc} &
        \min\{0,\frac{\bar\gamma_{i}(\bar\omega_{i} -
          \hat\omega_{i}^{\text{ref}}(k))}{
          \hat\omega_{i}^{\text{ref}}(k)-\bar\omega_{i}^{\text{thr}}}-v_{i}(k)\}
        & \text{if }\hat\omega_{i}^{\text{ref}}(k)>
        \bar\omega_{i}^{\text{thr}},
        \\
        & 0 & \hspace{1cm}\text{if }\underline\omega_{i}^{\text{thr}}\leqslant
        \hat\omega_{i}^{\text{ref}}(k)\leqslant \bar\omega_{i}^{\text{thr}},
        \\
        & \max\{0,\frac{\underline\gamma_{i}(\underline\omega_{i} -
          \hat\omega_{i}^{\text{ref}}(k))}{\underline\omega_{i}^{\text{thr}}
          - \hat\omega_{i}^{\text{ref}}(k)}-v_{i}(k) \} & \text{if
        }\hat\omega_{i}^{\text{ref}}(k)<
        \underline\omega_{i}^{\text{thr}},
      \end{array} \right. \;\hspace{1cm}\forall i\in\II{\omega},\;\forall k\in[0,N-1]_{\naturals},
    \\
    \hat
    u_{i}^{\text{ref}}(k)&\triangleq \sat(\hat
    u_{i}^{a}(k); \xi_{i},u_{i}^{\min},u_{i}^{\max}),\hspace{1cm}\forall i\in\II{\omega},\;\forall
    k\in[0,N-1]_{\naturals},\notag 
    \\
    \hat u_{i}^{\text{ref}}(k)&\triangleq0,\;\hspace{3.95cm}\forall i\in\mathcal{I}\backslash\II{\omega},\;\forall k\in[0,N-1]_{\naturals},\notag
    \\
    v_{i}(k)&\triangleq \sum_{j:j\rightarrow i}
    b_{ji}\hat\lambda_{ji}^{\text{ref}}(k)-\sum_{l:i\rightarrow
      l}b_{il}\hat \lambda_{il}^{\text{ref}}(k)+\hat
    p^{fcst}_{i}(k)-E_{i}\hat\omega_{i}^{\text{ref}}(k),\;\hspace{1cm}\forall
    i\in\II{\omega},\;\forall k\in[0,N-1]_{\naturals}.\notag 
  \end{align}
  \hrulefill
\vspace*{-2ex}
\end{figure*}
        
\begin{pf}
  From the definition of $(\hat
  \Lambda^{\text{ref}},\hat\Omega^{\text{ref}},\hat U^{\text{ref}})$
  one can easily see that it naturally satisfies
  constraints~\eqref{opti:nonlinear-1}-\eqref{opti:nonlinear-2b}
  and~\eqref{set:opti-nonlinear}.  We next show that the other
  constraints hold with each possible $\xi\in\{0,1\}^{|\II{u}|}$ by
  pointing out a specific $\hat B$ and $\hat \Gamma$ associated with
  $(\hat \Lambda^{\text{ref}},\hat\Omega^{\text{ref}},\hat
  U^{\text{ref}})$.  For any $i\in\II{u}$, if $\xi_{i}=0$, one can
  easily check that~\eqref{opti:nonlinear-4}-\eqref{opti:nonlinear-5}
  holds by the definition of $\hat u_{i}^{\text{ref}}$ with a trivial
  choice of $\beta_{i}(k)\equiv0$. Notice that since we assume that
  $\lambda_{i}$ is always $1$ if $\xi_{i}=0$, there always exists
  $\gamma_{i}(k)$ sufficiently large such
  that~\eqref{opti:nonlinear-3}-\eqref{opti:nonlinear-6} hold.

  If $\xi_{i}=1$ for some $i\in\II{u}$ instead, then one can have
  $\beta_{i}(k)$ sufficiently large to
  meet~\eqref{opti:nonlinear-4}-\eqref{opti:nonlinear-5}. Further if
  $\omega_{i,0}\nin [\underline \omega_{i},\bar\omega_{i}]$, resulting
  in $\lambda_{i}(\omega_{0},\xi_{i})=1$, then one can still choose
  $\gamma_{i}(k)$ sufficiently large so
  that~\eqref{opti:nonlinear-3}-\eqref{opti:nonlinear-6}
  hold. Finally, if $\omega_{i,0}\in [\underline
  \omega_{i},\bar\omega_{i}]$, then we show
  that~\eqref{opti:nonlinear-3}-\eqref{opti:nonlinear-6} also hold
  with a trivial choice of $\gamma_{i}(k)=0$ for every
  $k\in[1,N]_{\naturals}$. We first claim that there exists
  $c\in\real_{>}$ such that, for every $k\in[0,N-1]_{\naturals}$ and
  $i\in\mathcal{I}$,
  \begin{align}\label{ineq:omega-bound}
    |\hat\omega^{\text{ref}}_{i}(k+1)-\hat\omega^{\text{ref}}_{i}(k)|\leqslant
    cT.
  \end{align}
  Note that $\hat x^{\text{ref}} (k) \triangleq (\hat
  \lambda^{\text{ref}}(k),\hat\omega^{\text{ref}}(k))\in\real^{m+n}$,
  obtained by substituting $\hat u^{\text{ref}}$
  into~\eqref{opti:nonlinear-1}-\eqref{opti:nonlinear-2}, satisfies
  $\hat x^{\text{ref}}(k+1)=\hat x^{\text{ref}}(k)+Th(\hat
  x^{\text{ref}}(k),\hat p^{fcst}(k))$, which correspond to the Euler
  approximation of the continuous-time dynamics $ \dot{
    x}^{\text{ref}}(t)=h( x^{\text{ref}}(t),p^{fcst}_{t}(t))$.  Here,
  for simplicity, we omit the explicit expression of~$h$, but one can
  see that it is Lipschitz in its first component, and hence the
  solution of the continuous-time dynamics exists and is unique for
  any $t\geqslant 0$, and $\| x^{\text{ref}}(t)\|\leqslant r_{1}$ for
  sufficiently large $r_{1}\in\real_{>}$.
  By~\citep[Theorem~212A]{JCB:08}, there exists $c_{1}\in\real_{>}$
  such that
  \begin{align*}
    \|x^{\text{ref}}(\tau_{0}+kT)-\hat x^{\text{ref}}(k)\|\leqslant
    c_{1}T,\;\forall k\in[0,N-1]_{\naturals.}
  \end{align*}
  Further, the Lipschitz property of $h$ and the uniform boundedness
  of $x^{\text{ref}}(t)$ imply that there exists $r_{2}\in\real_{>}$
  such that $\|\dot x^{\text{ref}}(t)\|\leqslant r_{2}$ for any
  $t\geqslant \tau_{0}$. Therefore, it holds for all
  $k\in[0,N-1]_{\naturals}$ and all $i\in\mathcal{I}$ that
  \begin{align*}
    &\hspace{0.5cm}|\hat\omega^{\text{ref}}_{i}(k+1) -
    \hat\omega^{\text{ref}}_{i}(k)|\leqslant\|\hat
    x^{\text{ref}}(k+1)-\hat x^{\text{ref}}(k)\|
    \\
    &\leqslant \|\hat
    x^{\text{ref}}(k+1)-x^{\text{ref}}(\tau_{0}+(k+1)T)\|+\|\hat
    x^{\text{ref}}(k)- x^{\text{ref}}(\tau_{0}+kT)\|
    \\
    &\hspace{0.45cm}+ \|x^{\text{ref}}(\tau_{0}+(k+1)T)-
    x^{\text{ref}}(\tau_{0}+kT)\|
    \\
    &\leqslant2c_{1} T+\|\int_{kT}^{(k+1)T}\dot
    x^{\text{ref}}(\tau)\text{d}\tau\|
    \\
    &\leqslant 2c_{1}T+\sqrt{m+n}\int_{kT}^{(k+1)T}\|\dot
    x^{\text{ref}}(\tau)\|\text{d}\tau=(2c_{1}+r_{2}\sqrt{m+n})T.
  \end{align*}
  Hence,~\eqref{ineq:omega-bound} follows by letting $c\triangleq
  2c_{1}+r_{2}\sqrt{m+n}$.

  Now we first prove~\eqref{opti:nonlinear-3} holds for any
  $i\in\II{u}$ such that $M_{i}=0$.    For every $k\in[0,N]$, let 
  \begin{align*}
    \vartheta_{i}(k)&\triangleq
    v_{i}(k)+E_{i}\hat\omega_{i}^{\text{ref}}(k)
    \\
    &= \sum_{j:j\rightarrow i}
    b_{ji}\hat\lambda_{ji}^{\text{ref}}(k)-\sum_{l:i\rightarrow
      l}b_{il}\hat \lambda_{il}^{\text{ref}}(k)+\hat
    p^{fcst}_{i}(k),
  \end{align*}
  and note that $\vartheta_{i}(k)$ does not depend on
  $\hat\omega^{\text{ref}}(k)$.  Now, from the system dynamics, for
  each $i\in\II{u}$ such that $M_{i}=0$,
  \begin{align}\label{eqn:dae-ref}
    0=\vartheta_{i}(k)-E_{i}\hat\omega_{i}^{\text{ref}}(k)+\hat
    u^{\text{ref}}_{i}(k),
  \end{align}
  and one can check that~\eqref{eqn:dae-ref} possesses three possible
  solutions: a) $\hat\omega_{i}^{\text{ref}}(k)=\vartheta_{i}(k)\slash
  E_{i}$ with
  $\underline\omega_{i}\leqslant\vartheta_{i}(k)\slash E_{i}\leqslant\bar\omega_{i}$, b)  $\hat\omega_{i}^{\text{ref}}(k)=\bar\omega_{i}$ with
  $\vartheta_{i}(k)/E_{i} >  \bar\omega_{i} $, and c) 
$\hat\omega_{i}^{\text{ref}}(k)=\underline\omega_{i}$ with
  $\vartheta_{i}(k)/E_{i} <\underline \omega_{i}$.
%
%
Now it is easy to see that, depending on the value of
$\vartheta_{i}(k)/E_{i}$, the solution of
$\hat\omega_{i}^{\text{ref}}(k)$ is unique and always satisfies
$\underline\omega_{i}\leqslant\hat\omega_{i}^{\text{ref}}(k)\leqslant\bar\omega_{i}$.

At last,  we prove~\eqref{opti:nonlinear-3} holds for any $i\in\II{u}$ such that $M_{i}>0$   by induction, i.e.,
  for any $i\in\II{\omega}$, if $\hat\omega_{i}^{\text{ref}}(k)\in
  [\underline\omega_{i}, \bar\omega_{i}]$ for some
  $k\in[0,N-2]_{\naturals}$, then it also holds by replacing $k$ by
  $k+1$. Note that by~\eqref{ineq:omega-bound}, if
  $\hat\omega_{i}^{\text{ref}}(k)\in[\underline\omega_{i} +cT,
  \bar\omega_{i}-cT]$, then $\hat\omega_{i}^{\text{ref}}(k+1)\in
  [\underline\omega_{i}, \bar\omega_{i}]$. Therefore, we only need to
  consider the case when
  $\hat\omega_{i}^{\text{ref}}(k)\in(\bar\omega_{i}-cT,\bar\omega_{i}]$
  and
  $\hat\omega_{i}^{\text{ref}}(k)\in(\underline\omega_{i},\underline\omega_{i}+cT]$. For
  simplicity, we only prove the first case (the other holds
  similarly). Without loss of generality, we choose $T$ small enough
  so that $cT<\bar\omega_{i}-\bar\omega_{i}^{\text{thr}}$ for every
  $i\in\II{\omega}$, ensuring
  $\hat\omega_{i}^{\text{ref}}(k)>\bar\omega_{i}^{\text{thr}}$.
  From the system dynamics, one has
  $M_{i}\hat\omega_{i}^{\text{ref}}(k+1) =
  M_{i}\hat\omega_{i}^{\text{ref}}(k)+T\left(v_{i}(k)+ \hat
    u^{\text{ref}}_{i}(k)\right)$. Substituting~\eqref{eqn:ref-controller},
  one has
  \begin{align*}
    M_{i}\hat\omega_{i}^{\text{ref}}(k+1)&\leqslant
    M_{i}\hat\omega_{i}^{\text{ref}}(k) +
    T\frac{\bar\gamma_{i}(\bar\omega_{i} -
      \hat\omega_{i}^{\text{ref}}(k))}{\hat\omega_{i}^{\text{ref}}(k)
      - \bar\omega_{i}^{\text{thr}}}
    \\
    &\leqslant M_{i}\hat\omega_{i}^{\text{ref}}(k) +
    T\frac{\bar\gamma_{i}(\bar\omega_{i} -
      \hat\omega_{i}^{\text{ref}}(k))}{\bar\omega_{i} -
      cT-\bar\omega_{i}^{\text{thr}}}.
  \end{align*}
  By substituting $b(j) \triangleq \hat \omega_{i}^{\text{ref}}(j) -
  \bar\omega_{i}$ for $j=k$ and $k+1$ into the above inequality, it
  holds
  \begin{align*}
    M_{i}b(k+1)\leqslant 
    \left(M_{i}
      -\frac{T\bar\gamma_{i}}{\bar\omega_{i}-cT-\bar\omega_{i}^{\text{thr}}}\right)b(k).  
  \end{align*}
  Since $b(k)\leqslant 0$, let $\bar T$ be such that
  $M_{i}-\frac{\bar T\bar\gamma_{i}}{\bar\omega_{i}-c\bar
    T-\bar\omega_{i}^{\text{thr}}}>0$. Then, $b(j+1)\leqslant 0$ for
  $0<T\leqslant \bar T$, i.e., if $\hat\omega_{i}^{\text{ref}}(k)
  \leqslant \bar\omega_{i}$, then $\hat\omega_{i}^{\text{ref}}(k+1)
  \leqslant \bar\omega_{i}$, and the induction holds. \qed
\end{pf}

Notice that a small sampling length $T$ reduces the discretization gap
between $Q_{cont}$ and $Q_{disc}$, as well as guarantees the
qualification of $(\hat F^{\text{ref}},\hat\Omega^{\text{ref}},\hat
U^{\text{ref}})$ in Proposition~\ref{prop:ref-generation} as a
reference trajectory. On the other hand, the number of constraints
appearing in $Q_{cvx}$ grows linearly with respect to $1/T$. Hence,
there is a trade-off among discretization accuracy, reference
trajectory qualification, and computational complexity.

\section{From centralized to distributed closed-loop receding horizon 
  feedback}\label{sec:closed-loop-analysis}
        
In this section we design a feedback controller in a receding horizon
fashion by having the input at a given state $(\lambda(t),\omega(t))$
at time $t$ with a forecasted power injection $p_{t}^{fcst}$ be the
first step of the optimal control input trajectory of
$Q_{cvx}(\mathcal{G},\II{u},\II{\omega},\hat
P^{fcst},\lambda(t),\omega(t),t)$. We first consider a centralized
implementation, where a single operator gathers global state
information, computes the control law, and broadcasts it. Building on
it, we propose a distributed strategy, where several independent
operators are responsible for computing control signals within its own
region using only regional information.

%

\subsection{Centralized control with stability and frequency
  invariance}\label{subsection:c-MPC}
        
Formally, at time $t$, the centralized controller measures the current
output $(f(t),\omega(t))$ and forecasts a power injection profile
$p^{fcst}_{t}(\tau)$ with $\tau\in[t,t+\tilde t]$ as well as its
corresponding discretization $\hat P^{fcst}$,
cf.~\eqref{eqn:p-G-graph}.
Let $(\hat \Lambda^{*}_{cvx},\hat\Omega^{*}_{cvx},\hat U^{*}_{cvx})$ be the
optimal solution of
$Q_{cvx}(\mathcal{G},\II{u},\II{\omega},\hat
P^{fcst},f(t),\omega(t),t)$. The centralized control law is then given
by
\begin{align}\label{eqn:control-input-mpc}
  u(x(t),p^{fcst}_{t})\triangleq \hat u^{*}_{cvx}(0),
\end{align}
where $ u^{*}_{cvx}(0)$ is the first column of $\hat U^{*}_{cvx}$.
The next result states that the controller is able to stabilize the
system without changing its open-loop equilibrium point, and, at the
same time, guarantees safe frequency region invariance and
attractivity.

\begin{theorem}\longthmtitle{Centralized control with stability and
    frequency constraints}\label{thm:cen-control}
  Under Assumption~\ref{assumption:forecast-injection} and for any
  initial state $(\lambda(0),\omega(0))$, the closed-loop
  system~\eqref{eqn:compact-form} with
  controller~\eqref{eqn:control-input-mpc} and sufficiently small
  sampling length $T$ satisfies:
  \begin{enumerate}
  \item\label{item:zero-input} For any $i\in\II{u}$ with any
    $\xi_{i}\in\{0,1\}$ and any $t\in\real_{\geqslant }$,
    $u_{i}(x(t),p^{fcst}_{t})=0$ if
    $\omega_{i}(t)\in(\underline\omega_{i}^{\text{thr}},\bar\omega_{i}^{\text{thr}})$;
  \item\label{item:frequency-invariance} For any $i\in\II{\omega}$
    with $\xi_{i}=1$, if $\omega_{i}(0)\in[\underline\omega_{i},
    \bar\omega_{i}]$, then $\omega_{i}(t)\in[\underline\omega_{i},
    \bar\omega_{i}]$ for any $t\geqslant 0$.
  \end{enumerate}
  Furthermore, if in addition Assumption~\ref{assumption:TL-power} and
  condition~\eqref{ineq:sufficient-eq} hold, and
  $(\lambda(0),\omega(0))\in\Phi (r)$ with some $0\leqslant r<\bar r$,
  then:
  \begin{enumerate}
    \addtocounter{enumi}{2}
  \item\label{item:convergence} For any $\xi\in\{0,1\}^{|\II{u}|}$,
    $(\lambda^{\infty},\omega^{\infty}\ones_{n})$ is locally
    asymptotically stable, $(\lambda(t),\omega(t))\in\Phi (r)$ for
    every $t\geqslant 0$, and
    $(\lambda(t),\omega(t))\rightarrow(\lambda^{\infty},\omega^{\infty}\ones_{n})$;
  \item\label{item:finite-convergence} For any $i\in\II{u}$ with any
    $\xi_{i}\in\{0,1\}$, $u_{i}(x(t),p^{fcst}_{t})$ converges to $0$
    in finite time;
  \item\label{item:frequency-attractive} For any $i\in\II{\omega}$
    with $\xi_{i}=1$, if $\omega_{i}(0)\nin[\underline\omega_{i},
    \bar\omega_{i}]$, then there exists a finite $t_{1}$ such that
    $\omega_{i}(t)\in[\underline\omega_{i}, \bar\omega_{i}]$ for any
    $t\geqslant t_{1}$.
  \end{enumerate}
\end{theorem}
\begin{pf}
  We first show that $u$ is well-defined by proving that $ \hat
  u^{*}_{cvx}(0)$ exists and is unique. Notice that $(\hat
  \Lambda^{\text{ref}},\hat\Omega^{\text{ref}},\hat U^{\text{ref}})$
  defined in Proposition~\ref{prop:ref-generation} always qualifies as
  a reference trajectory for sufficiently small $T$. Hence the
  feasible set of $Q_{cvx}$ is non-empty, and thus there exists at
  least one optimal solution.  Uniqueness follows from the strict
  convexity of the objective function.
  For~\ref{item:zero-input}, note that in
  $Q_{cvx}(\mathcal{G},\II{u},\II{\omega},\hat
  P^{fcst},\lambda(t),\omega(t),t)$, if
  $\omega_{i}(t)\in(\underline\omega_{i}^{\text{thr}},\bar\omega_{i}^{\text{thr}})$
  for some $i\in\II{u}$, then by~\eqref{set:opti-linear-phi} and the
  fact that $\hat\omega_{i}^{\text{ref}}(0)=\omega_{i}(t)$, one has
  $\hat u_{i,cvx}^{*}(0)=0$, and hence the statement follows
  by~\eqref{eqn:control-input-mpc}.
          
  The statement in~\ref{item:frequency-invariance} is
  equivalent~\citep[Lemma~4.3]{YZ-JC:19-auto} to
  \begin{subequations}\label{ineq:invariance-condition-unsymmetric-1}
    \begin{align}
      \dot\omega_{i}(t)\leqslant 0,\ \text{if }
      \omega_{i}(t)=\bar\omega_{i},\label{ineq:invariance-condition-unsymmetric-1a}
      \\
      \dot\omega_{i}(t)\geqslant 0,\ \text{if
      }\omega_{i}(t)=\underline
      \omega_{i}.\label{ineq:invariance-condition-unsymmetric-1b}
    \end{align}
  \end{subequations}
  For simplicity, here we only
  prove~\eqref{ineq:invariance-condition-unsymmetric-1a}. Since
  $(\hat\Lambda^{*}_{cvx},\hat\Omega^{*}_{cvx}, \hat U^{*}_{cvx})$ is
  feasible for $Q_{cvx}(\mathcal{G},\II{u},\II{\omega},\hat
  P^{fcst}$, $\lambda(t),\omega(t),t)$, it satisfies
  constraint~\eqref{opti:nonlinear}.  Extracting the $i$th equation
  with $k=1$ from~\eqref{opti:nonlinear-1}, it holds
  \begin{align}\label{eqn:single-estimate-dynamics}
    &M_{i}\hat\omega_{i,cvx}^{*}(1) = M_{i}\hat\omega_{i,cvx}^{*}(0) +
    T\big\{-E_{i}\hat\omega_{i,cvx}^{*}(0)-[D^{T}Y_{b}]_{i}
    \hat\lambda_{cvx}^{*}(0)\notag
    \\
    &\hspace{4cm}+\hat p^{fcst}_{i}(0)+\hat u_{i,cvx}^{*}(0)\big\}.
  \end{align}
  Note first, by~\eqref{opti:nonlinear-2}, $\hat
  \lambda_{cvx}^{*}(0)=\sin\lambda(t)$ and
  $\hat\omega_{i,cvx}^{*}(0)=\omega_{i}(t)$; secondly,
  $u_{i}(x(t),p^{fcst}_{t})=\hat u_{i,cvx}^{*}(0)$; thirdly, $\hat
  p^{fcst}_{i}(0)\triangleq p^{fcst}_{i,t}(t)$, which by assumption
  equals
  $p_{i}(t)$; 
  fourthly, by~\eqref{opti:nonlinear-3},
  $\hat\omega_{i,cvx}^{*}(1)\leqslant \bar\omega_{i}$. These four
  facts imply that, when $\omega_{i}(t)=\bar\omega_{i}$,
  \begin{align}\label{ineq:nagumo-frequency}
    -E_{i}\bar\omega_{i}(t)-[D^{T}]_{i}Y_{b} \sin\lambda(t)+
    p_{i}(t)+u_{i}(x(t),p^{fcst}_{t})\leqslant 0.
  \end{align}
  From~\eqref{eqn:compact-form-2}, one sees
  that~\eqref{ineq:nagumo-frequency} is
  exactly~\eqref{ineq:invariance-condition-unsymmetric-1a}, concluding
  our reasoning.
  
  To prove statement~\ref{item:convergence}, since $(\hat
  \Lambda^{*}_{cvx},\hat\Omega^{*}_{cvx}, \hat
  U^{*}_{cvx})\in\Phi_{cvx}$, by Lemma~\ref{lemma:convexificaiton},
  one has $(\hat \Lambda^{*}_{cvx},\hat\Omega^{*}_{cvx}, \hat
  U^{*}_{cvx})\in\Phi_{disc}$, which further implies that for every
  $i\in\II{u},$
  \begin{align*}
    \hat\omega_{i,cvx}^{*}(0)\hat u_{i,cvx}^{*}(0)\leqslant 0, \ \text{if
    }\hat\omega_{i,cvx}^{*}(0)\nin
    (\underline\omega_{i}^{\text{thr}},\bar\omega_{i}^{\text{thr}}),
    \\
    \hat u_{i,cvx}^{*}(0)=0, \ \text{if }\hat\omega_{i,cvx}^{*}(0)\in
    (\underline\omega_{i}^{\text{thr}},\bar\omega_{i}^{\text{thr}}).
  \end{align*}
  Since $\hat\omega_{i,cvx}^{*}(0)=\omega_{i}(t)$, together with the
  definition of controller~\eqref{eqn:control-input-mpc} and
  Lemma~\ref{prop:robust-as}, it holds that the closed-loop system is
  asymptotically stable.

  To prove statement~\ref{item:finite-convergence}, since we have
  already shown the converge of $(\lambda(t),\omega(t))$, it holds
  that for arbitrarily small $\delta\in\real_{>}$, there exists
  $\tilde t\in\real_{\geqslant }$ such that
  $|\omega_{i}(t)-\omega^{\infty}|< \delta$ for any $i\in\II{u}$ at
  any $t\geqslant \tilde t$.  Let
  $\delta\triangleq\min_{i\in\II{u}}\{\min(\bar\omega^{\text{thr}}_{i}
  - \omega^{\infty},\omega^{\infty} -
  \underline\omega_{i}^{\text{thr}})\}>0$. Now consider any
  $t\geqslant \tilde t$, one has
  $\omega_{i}(t)\in(\underline\omega_{i}^{\text{thr}},\bar\omega_{i}^{\text{thr}})$,
  which, by statement~\ref{item:zero-input}, implies
  $u_{i}(x(t),p_{t}^{fcst})=0$.

  Finally, to prove statement~\ref{item:frequency-attractive},
  by~\ref{item:convergence}, since every $\omega_{i}$ ultimately
  converges to $\omega^{\infty}$, it must first enter
  $[\underline\omega_{i},\bar\omega_{i}]$, which,
  by~\ref{item:frequency-invariance}, cannot leave the safe region
  afterwards. \qed
\end{pf}

\begin{remark}\longthmtitle{Independence of stability on prediction
    model}\label{rmk:stability-linear-MPC}
  {\rm Since the prediction model~\eqref{opti:nonlinear-1} is
    linearized and discretized based on the true nonlinear
    dynamics~\eqref{eqn:compact-form}, it naturally brings state
    prediction error into the feedback control design; however, this
    does not jeopardize closed-loop asymptotic stability because we
    impose the stability constraint~\eqref{set:opti-linear-phi} which
    is independent of the prediction model. That being said, the model
    mismatch could lead to loss of optimality.  \oprocend}
\end{remark}

Theorem~\ref{thm:cen-control}\ref{item:frequency-attractive} states
the finite-time recovery of frequency property to the safe interval
within time $t_{1}$. However, it is challenging to derive an
analytical expression for how depends on the design parameters (e.g.,
$c_{i}$, $d_{i}$, $e_{i}$ and $\gamma_{i}$). A basic observation is
that, since $e_{i}$ represents the penalty coefficient of the
predicted frequency violation in the objective function in
$(Q_{cvx})$, larger $e_{i}$ yields faster convergence from outside the
safe interval, leading to smaller~$t_{1}$.

Note that to compute the centralized control signal
in~\eqref{eqn:control-input-mpc}, the operator should complete the
following steps at every time: a)~collect state information and
forecast power injection of the entire network, b)~determine the
optimal trajectory $\hat U_{cvx}^{*}$ by solving $Q_{cvx}$, and
c)~broadcast the control signals to the corresponding controllers.
The time to complete any of these three steps grows with the size of
the network, which motivates the developments of our next section.

\subsection{Distributed control using regional information}
Here we describe our approach to design a distributed control strategy
that takes advantage of cooperation to optimize control effort while
ensuring stability and frequency invariance.  The idea is to divide
the power network into regions, and have each controller make
decisions based on the state and power injection prediction
information within its region. The network partition relies on the
following assumption.

\begin{assumption}\longthmtitle{Controlled nodes in induced
    subgraphs}\label{assumption:subgraph-node}
  Let $\mathcal{G}_{\beta}=(\mathcal{I}_{\beta},\mathcal{E}_{\beta}),\
  \beta\in[1,d]_{\naturals}$ be induced subgraphs of $\mathcal{G}$
  (i.e., $\mathcal{I}_{\beta}\subseteq\mathcal{I}$,
  $\mathcal{E}_{\beta}\subseteq\mathcal{E}$, and
  $(i,j)\in\mathcal{E}_{\beta}$ if $(i,j)\in\mathcal{E}$ with
  $i,j\in\mathcal{I}_{\beta}$). We assume that each controlled node is
  contained in one and only one region, i.e.,
  \begin{subequations}\label{sube:assu-subgraph}
    \begin{align}
      &\II{u}\subseteq\bigcup_{\beta=1}^{d}\mathcal{I}_{\beta},\label{sube:assu-subgraph-1} 
      \\
     &\mathcal{I}_{\alpha}\bigcap\mathcal{I}_{\beta} \bigcap \II{u}=
  \emptyset,\ \forall \alpha,\beta\in [1,d]_{\naturals}
      \text{ with } \alpha\neq\beta.\label{sube:assu-subgraph-2}
    \end{align}
  \end{subequations}
\end{assumption}

The induced subgraphs represent the regions of the network.  Our
distributed control strategy consists of implementing the centralized
control for every induced subgraph $\mathcal{G}_{\beta}$, where for
every line $(i,j)\in\mathcal{E}_{\beta}' \subseteq
\mathcal{I}_{\beta}\times(\mathcal{I}\backslash\mathcal{I}_{\beta})$
connecting $\mathcal{G}_{\beta}$ and the rest of the network, we treat
its power flow $f_{ij}(\tau)$ as an external power injection whose
forecasted value is a constant equaling its current value $f_{ij}(t)$
for $\tau\in[t,t+\tilde t]$.  Formally, 
\begin{align}\label{eqn:pfcst-f}
  p_{t,\beta,i}^{fcst,f}(\tau)\triangleq\sum_{\substack{j\rightarrow
      i\\(i,j)\in\mathcal{E}'}}f_{ij}(t)-\sum_{\substack{i\rightarrow
      j\\(i,j)\in\mathcal{E}'}}f_{ij}(t),\;\forall \tau\in [t,t+\tilde
  t],
\end{align}
as the forecasted (starting from the current time $t$) power flow from
transmission lines in $\mathcal{E}_{\beta}'$ injecting into node
$i\in\mathcal{I}_{\beta}$. Let $p^{fcst,f}_{t,\beta}:[t,t+\tilde
t]\rightarrow\real^{|\mathcal{I}_{\beta}|}$ be the collection of all
such $p^{f}_{t,\beta,i}$'s with $i\in\mathcal{I}_{\beta}$. Also, let
$p^{fcst}_{t,\beta}:[t,t+\tilde
t]\rightarrow\real^{|\mathcal{I}_{\beta}|}$ be the collection of all
$p^{fcst}_{t,i}$'s with $i\in\mathcal{I}_{\beta}$, and denote $
p_{t,\beta}^{fcst,o}\triangleq
p^{fcst,f}_{t,\beta}+p^{fcst}_{t,\beta}$ as the overall forecasted
power injection for $\mathcal{G}_{\beta}$. Denote $\hat
P^{fcst,o}_{\beta}$ as its discretization.  Define
$\II{u}_{\beta}\triangleq\II{u}\bigcap\mathcal{I}_{\beta}$
(resp. $\II{\omega}_{\beta}\triangleq\II{\omega}\bigcap\mathcal{I}_{\beta}$)
as the collection of nodes within $\mathcal{G}_{\beta}$ with available
controllers (resp. with frequency constraints). Let
$(f_{\beta},\omega_{\beta})\in\real^{|\mathcal{I}_{\beta}|+|\mathcal{E}_{\beta}|}$
be the collection of states within $\mathcal{G}_{\beta}$.

Similarly to~\eqref{eqn:control-input-mpc}, let $(\hat
\Lambda^{*}_{cvx,\beta},\hat\Omega^{*}_{cvx,\beta},\hat U^{*}_{cvx,\beta})$
be the optimal solution of
$Q_{cvx}(\mathcal{G}_{\beta},\II{u}_{\beta},\mathfrak{G_{\beta}},\hat
P^{fcst,o}_{\beta},f_{\beta}(t),\omega_{\beta}(t),t)$. The control law
is given by
\begin{align}\label{eqn:control-input-mpc-dis}
  u_{i}(x(t),p^{fcst}_{t})\triangleq \hat
  u^{*}_{i,cvx,\beta}(0),\;\forall i\in\II{u},
\end{align}
where $u^{*}_{i,cvx,\beta}(0)$ is the $i$th entry of
$u^{*}_{cvx,\beta}(0)$ (the first column of $\hat
U^{*}_{cvx,\beta}$). 

To implement the controller~\eqref{eqn:control-input-mpc-dis} in a
distributed fashion, each region $\mathcal{G}_{\beta}$ with
$\beta\in[1,d]_{\naturals}$, independently of the rest, measures
system information within itself and power flows across its
boundary. After this, each region solves its own optimization problem
$Q_{cvx}(\mathcal{G}_{\beta},\II{u}_{\beta},\mathfrak{G_{\beta}},\hat
P^{fcst,o}_{\beta},f_{\beta}(t),\omega_{\beta}(t),t)$ and broadcasts
the solution $\hat u^{*}_{i,cvx,\beta}(0)$ to each node $i\in\II{u}$
within~$\mathcal{G}_{\beta}$.  The next result details the properties
of this strategy.

\begin{proposition}\longthmtitle{Distributed control with stability
    and frequency constraints}\label{prop:dis-control}
  Given power injection $p$ and any initial state
  $(f(0),\omega(0))\in\Gamma$, under
  Assumptions~\ref{assumption:forecast-injection}
  and~\ref{assumption:subgraph-node} with sufficiently small sampling
  length $T$, the following statements hold for the closed-loop
  system~\eqref{eqn:compact-form} under
  controller~\eqref{eqn:control-input-mpc-dis}:
  \begin{enumerate}
  \item\label{item:zero-input-dis} For any $i\in\II{u}$ with any
    $\xi_{i}\in\{0,1\}$ and any $t\in\real_{\geqslant }$,
    $u_{i}(x(t),p^{fcst}_{t})=0$ if
    $\omega_{i}(t)\in(\underline\omega_{i}^{\text{thr}},\bar\omega_{i}^{\text{thr}})$;
  \item\label{item:frequency-invariance-dis} For any
    $i\in\II{\omega}$ with $\xi_{i}=1$, if
    $\omega_{i}(0)\in[\underline\omega_{i}, \bar\omega_{i}]$, then
    $\omega_{i}(t)\in[\underline\omega_{i}, \bar\omega_{i}]$ for any
    $t\geqslant 0$.
  \end{enumerate}
  Furthermore, if in addition Assumption~\ref{assumption:TL-power} and
  condition~\eqref{ineq:sufficient-eq} hold, and
  $(\lambda(0),\omega(0))\in\Phi (r)$ with some $0\leqslant r<\bar r$,
  then:
  \begin{enumerate}
    \addtocounter{enumi}{2}
  \item\label{item:convergence-dis}
    $(\lambda^{\infty},\omega^{\infty}\ones_{n})$ is locally
    asymptotically stable, $(\lambda(t),\omega(t))\in\Phi (r)$ for
    every $t\geqslant 0$, and
    $(\lambda(t),\omega(t))\rightarrow(\lambda^{\infty},\omega^{\infty}\ones_{n})$;
  \item\label{item:finite-convergence-dis} For any $i\in\II{u}$ with
    any $\xi_{i}\in\{0,1\}$, $u_{i}(x(t),p^{fcst}_{t})$ converges to
    $0$ within a finite time;
  \item\label{item:frequency-attractive-dis} For any $i\in\II{\omega}$
    with $\xi_{i}=1$, if $\omega_{i}(0)\nin[\underline\omega_{i},
    \bar\omega_{i}]$, then there exists a finite $t_{1}$ such that
    $\omega_{i}(t) \in [\underline\omega_{i}, \bar\omega_{i}]$ for any
    $t\geqslant t_{1}$.
  \end{enumerate}
\end{proposition}
\begin{pf}
  First notice that each $u_{i}$ is well-defined, as by
  Assumption~\ref{assumption:subgraph-node}, for every $i\in\II{u}$,
  $u_{i}$ is assigned to one and only one subgraph, and hence $\hat
  u^{*}_{i,cvx,\beta}(0)$ is determined uniquely by a single
  $Q_{cvx}(\mathcal{G}_{\beta},\II{u}_{\beta},\mathfrak{G_{\beta}},\hat
  P^{fcst}_{\beta},\lambda_{\beta}(t),\omega_{\beta}(t),t)$. The proofs of
  all statements follow similar arguments as the ones in
  Theorem~\ref{thm:cen-control}. For
  statement~\ref{item:frequency-invariance-dis}, similar to the way we
  have~\eqref{ineq:nagumo-frequency}, it holds that when
  $\omega_{i}(t)=\bar\omega_{i}$,
  $-E_{i}\bar\omega_{i}(t)-[D^{T}_{\beta}]_{i}
  f_{\beta}(t)+p_{t,\beta,i}^{fcst,f}(t)+p_{i}(t)+u_{i}(x(t),p^{fcst}_{t})\leqslant
  0$, where $D_{\beta}$ is the incidence matrix for
  $\mathcal{G}_{\beta}$. Notice that this inequality is equivalent
  to~\eqref{ineq:nagumo-frequency} as $[D^{T}_{\beta}]_{i}
  f_{\beta}(t)+p_{t,\beta,i}^{fcst,f}(t)=-[D^{T}]_{i} f(t)$
  by~\eqref{eqn:pfcst-f}, implying frequency invariance. \qed
\end{pf}

Although the statements in Theorem~\ref{thm:cen-control} and
Proposition~\ref{prop:dis-control} are similar, their corresponding
controllers~\eqref{eqn:control-input-mpc}
and~\eqref{eqn:control-input-mpc-dis} are in general not
equivalent. To see this point, note that each $u_{i}$ with
$i\in\II{u}$ defined in~\eqref{eqn:control-input-mpc} is a function of
the entire system information; however, each $u_{i}$
in~\eqref{eqn:control-input-mpc-dis} only depends on local information
within the region node $i$ belongs to. Such a local dependence allows
each region to independently compute its own optimization problem,
which is of a size significantly smaller than the global optimization.
The regional partition, however, induces less cooperation among
different regions (this is illustrated in the simulations below).

\section{Simulations}

\begin{figure}[htb]
  \centering%
  \includegraphics[width=.8\linewidth]{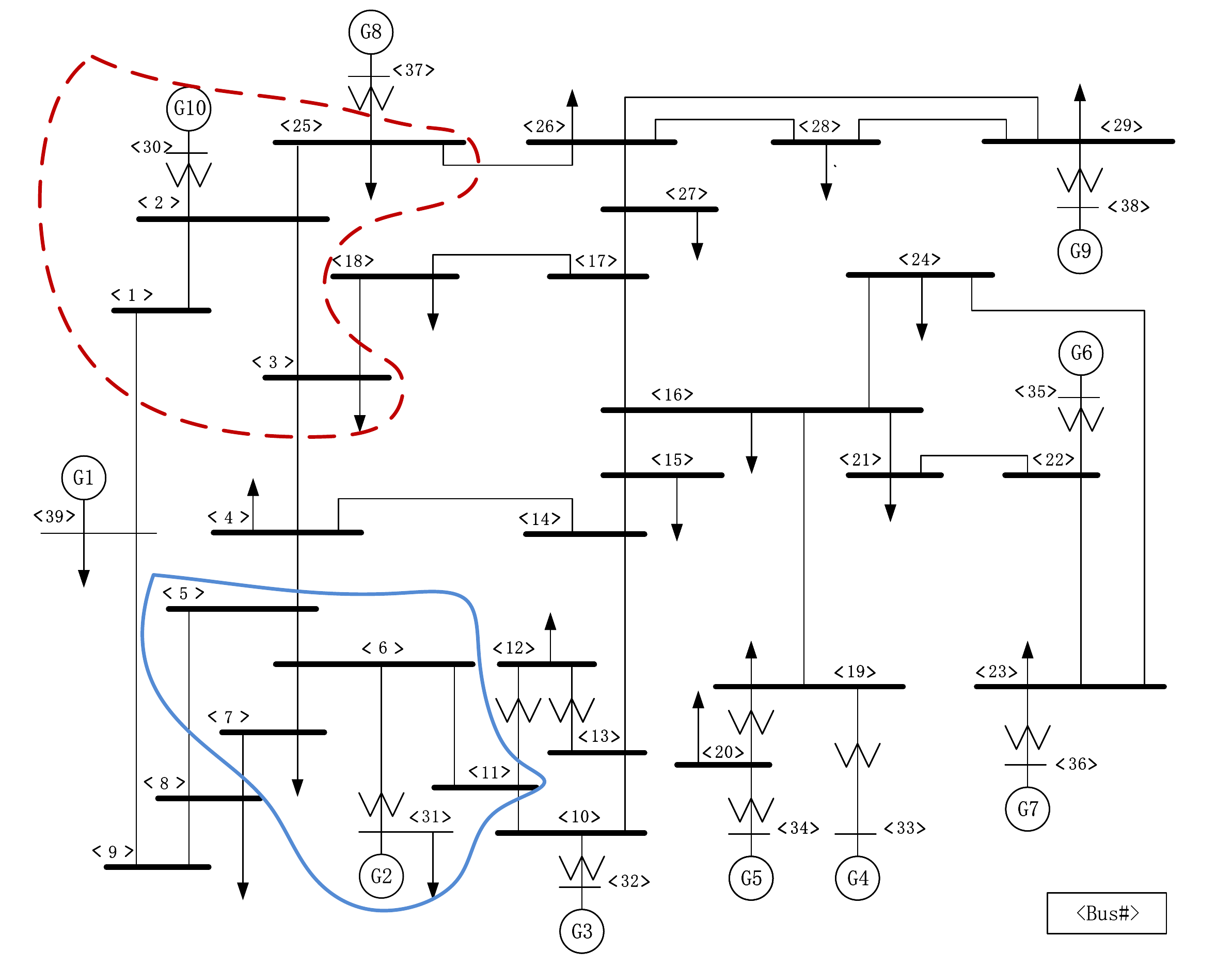}
  \caption{IEEE 39-bus power network.}\label{fig:IEEE39bus}
\end{figure}

We first illustrate the performance of the distributed controller in
the IEEE 39-bus power network displayed in Fig.~\ref{fig:IEEE39bus}.
We take the values of initial power injection $p_{i}(0)$, susceptance
$b_{ij}$, and rotational inertia $M_{i}$ from the Power System
Toolbox~\citep{KWC-JC-GR:09}, where nodes 30 to 39 (as generators)
possess strictly positive inertia, and the remaining 29 nodes have no
inertia.  The damping parameter is $E_{i}=1$ for all buses. 
The initial state $( \lambda(0),\omega(0))$ is chosen to be the
equilibrium with respect to the initial power injections. Let
$\II{\omega}=\{30,31\}$ be the two generators with transient frequency
requirements. As shown in Fig.~\ref{fig:IEEE39bus}, we assign each of
them a region containing its 2-hop neighbors.  Let
$\II{u}=\{3,7,25,30,31\}$ be the collection of nodal indexes with
controllers. Notice that Assumption~\ref{assumption:subgraph-node}
holds in this scenario.  To set up the optimization problem $Q_{cvx}$
so as to define our controller~\eqref{eqn:control-input-mpc-dis}, for
every $i\in\II{u}$, we set $\bar\gamma_{i}=\underline\gamma_{i}=1$
required in~\eqref{eqn:ref-controller}, $c_{i}=2$ if $i\in\II{\omega}$
and $c_{i}=1$ if $i\in\II{u}\backslash\II{\omega}$, $T=0.001s$. As a
trade-off between computation complexity and prediction horizon, we
select $N=150$ so that $\tilde t=0.15$s. For simplicity, for every
$i\in\II{u}$, let $\xi_{i}=1$ and $d_{i}=0$, i.e., we impose neither
hard nor soft constraints on the control signal amplitude, and
therefore, there is no need to specify $u_{i}^{\min}$ and
$u_{i}^{\max}$.  For every $i\in\II{\omega}$, let $e_{i}=500$,
$\bar\omega_{i}=-\underline\omega_{i}=0.2$Hz and
$\bar\omega_{i}^{\text{thr}}=-\underline\omega_{i}^{\text{thr}}=0.1$Hz.
The nominal frequency is 60Hz, and hence the safe frequency region is
$[59.8\text{Hz},\ 60.2\text{Hz}]$. {We take
  $p^{fcst}_{t}(\tau)=(1+\tau-t)p(\tau)$ for every $\tau\in[t,t+\tilde
  t]$, that is, the forecasted power injection error
  $p^{fcst}_{t}(\tau)-p(\tau)$ satisfies
  Assumption~\ref{assumption:forecast-injection}, and grows linearly
  in time.}

\begin{figure}[tbh!]
  \centering
  \subfigure[\label{frequency-response-no-control-generator}]{\includegraphics[width=.32\linewidth]{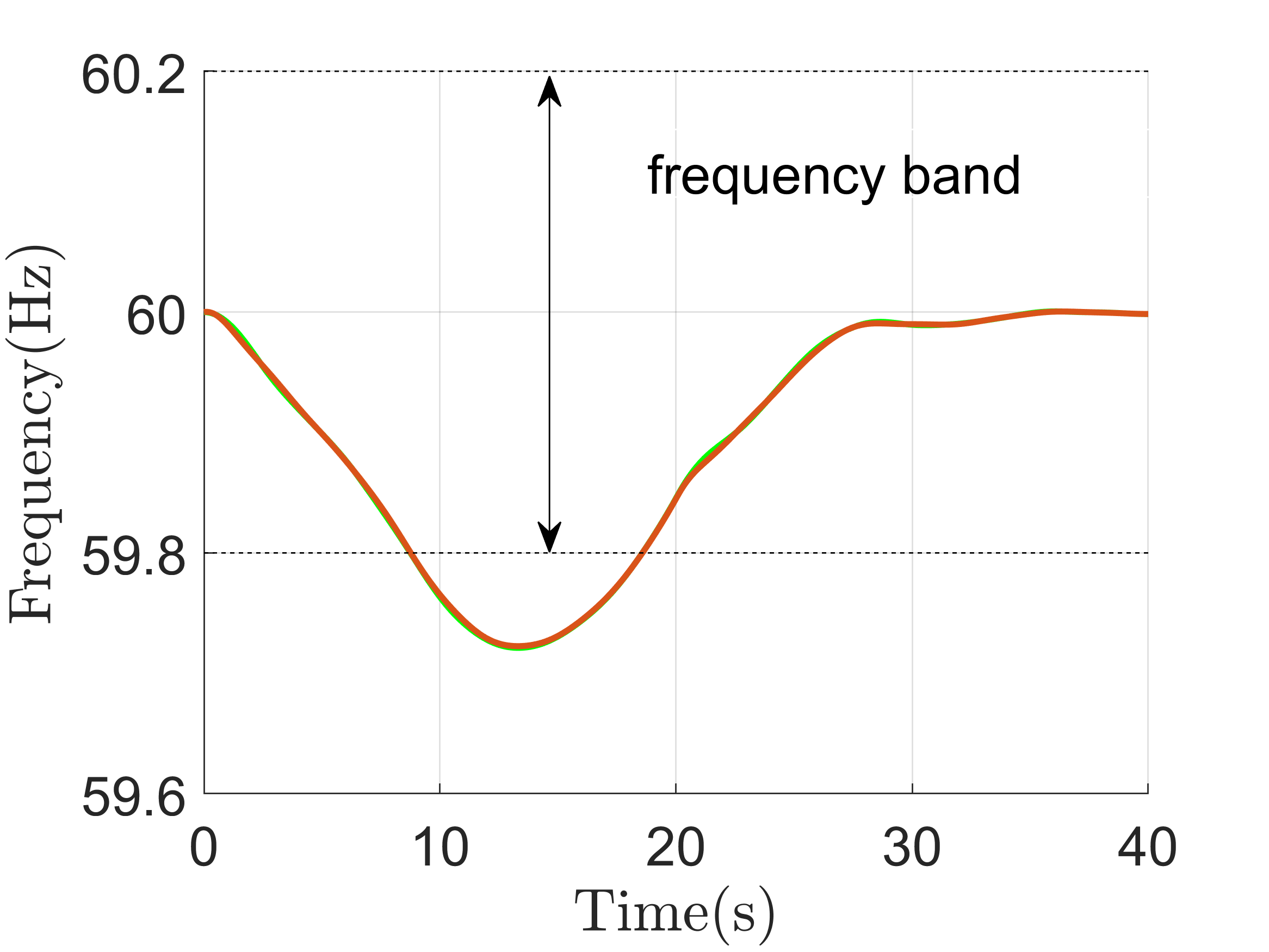}}
  \subfigure[\label{frequency-response-with-control-generator-region1}]{\includegraphics[width=.32\linewidth]{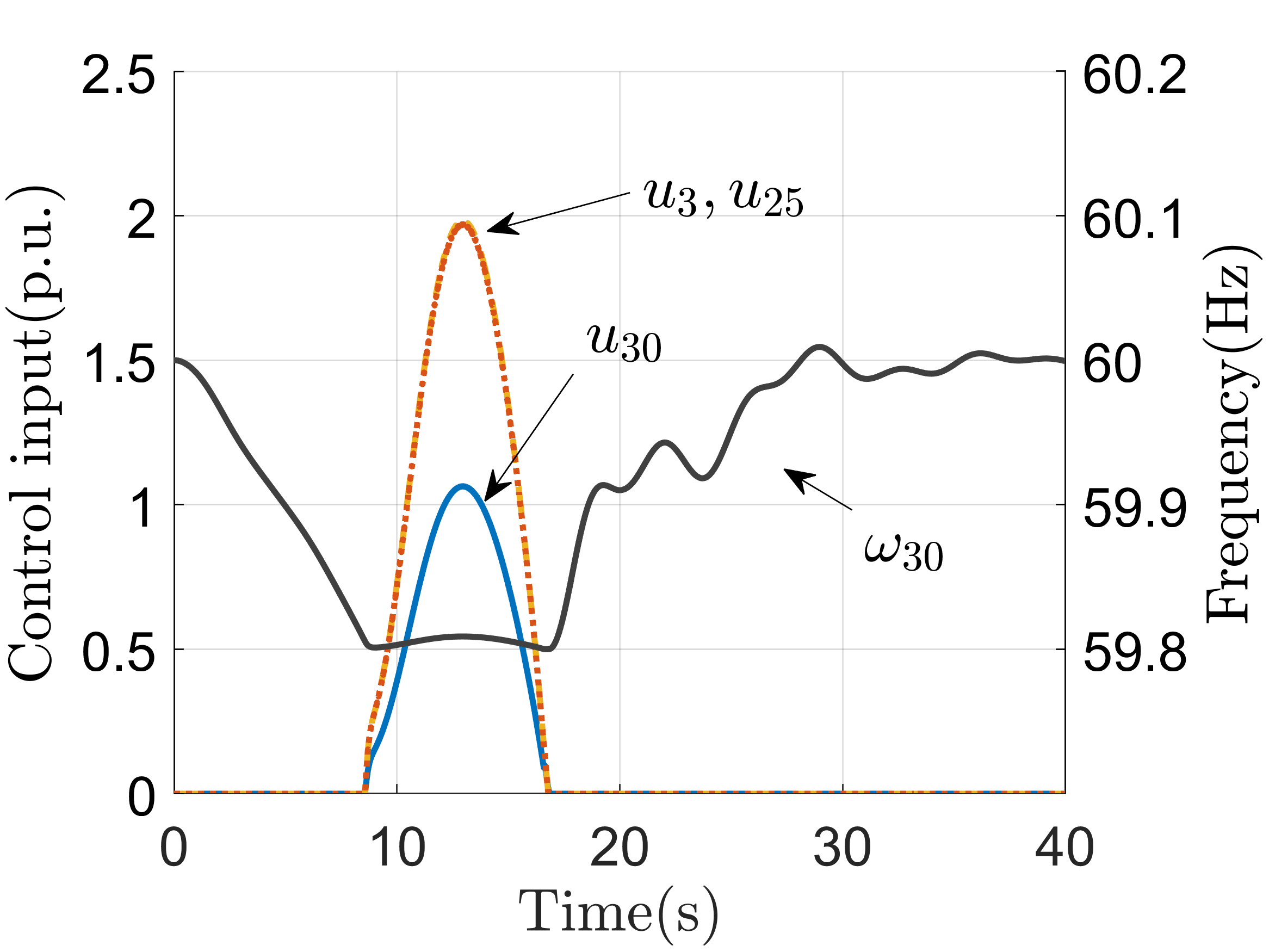}}
  \subfigure[\label{frequency-response-with-control-generator-region2}]{\includegraphics[width=.32\linewidth]{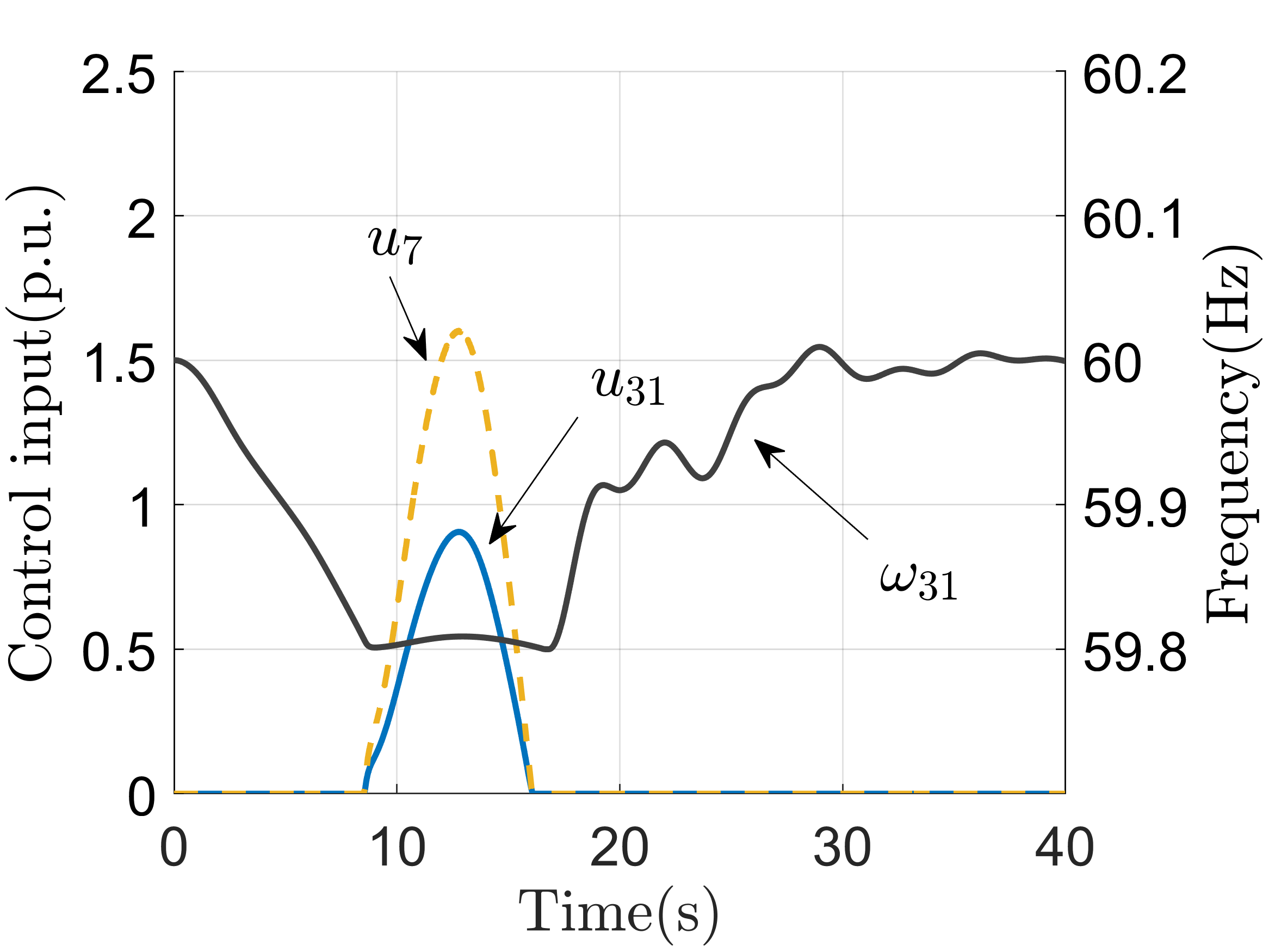}}
  \caption{Plot~\subref{frequency-response-no-control-generator} shows
    the frequency trajectories of generators 30 and 31 without the 
    controller, going beyond the lower safe frequency bound. With the centralized
    controller,
    plot~\subref{frequency-response-with-control-generator-region1}
    and \subref{frequency-response-with-control-generator-region2}
    show the trajectories of the control inputs and frequency within
    each region. }\label{fig:trajectories}
\end{figure}

We show that the proposed controller is able to maintain the targeted
generator frequencies within the safe region, provided that these
frequencies are initially in the safe region. We perturb all
non-generator nodes by a sinusoidal power injection whose magnitude is
proportional to the corresponding node's initial power
injection. Specifically, for every $i\in\{1,2,\cdots,29\}$, let
$p_{i}(t)=(1+\delta(t))p_{i}(0)$, where $\delta(t)= 0.25\sin(\pi
t\slash 20 )$ for $t<20$, and $\delta(t)=0$ for $t\geqslant 20$.
For $i\in\{30,31,\cdots,39\}$, let $p_{i}(t)\equiv p_{i}(0)$.
Fig.~\ref{fig:trajectories}\subref{frequency-response-no-control-generator}
shows the open-loop frequency responses of the two generators without
the controller. One can see that both trajectories exceed the lower
bound around~8s. With the distributed control,
Fig.~\ref{frequency-response-with-control-generator-region1}
and~\subref{frequency-response-with-control-generator-region2} show
the frequency and control input responses in the left-top region and
left-bottom region, resp.  Both frequency responses stay within the
safe bound all the time and converge to $60$Hz. Also, all control
signals vanish to 0 within~20s.  In
Fig.~\ref{frequency-response-with-control-generator-region1}, since we
assign a higher cost weight on $u_{30}$, and the same weight on
$u_{25}$ and $u_{3}$, the latter two have almost overlapping trajectory with
magnitude higher than the first one.  On the other hand, notice that
for every $i\in\II{\omega}$, $u_{i}$ is always 0, while $\omega_{i}$
is above the lower frequency threshold denoted by the dashed line. All
these observations are in agreement with the result of
Proposition~\ref{prop:dis-control}\ref{item:zero-input-dis}-\ref{item:finite-convergence-dis}
(even though the time-varying power injection used here does not
satisfy Assumption~\ref{assumption:TL-power}).

\begin{figure}[htb]
  \centering%
  \includegraphics[width=.65\linewidth]{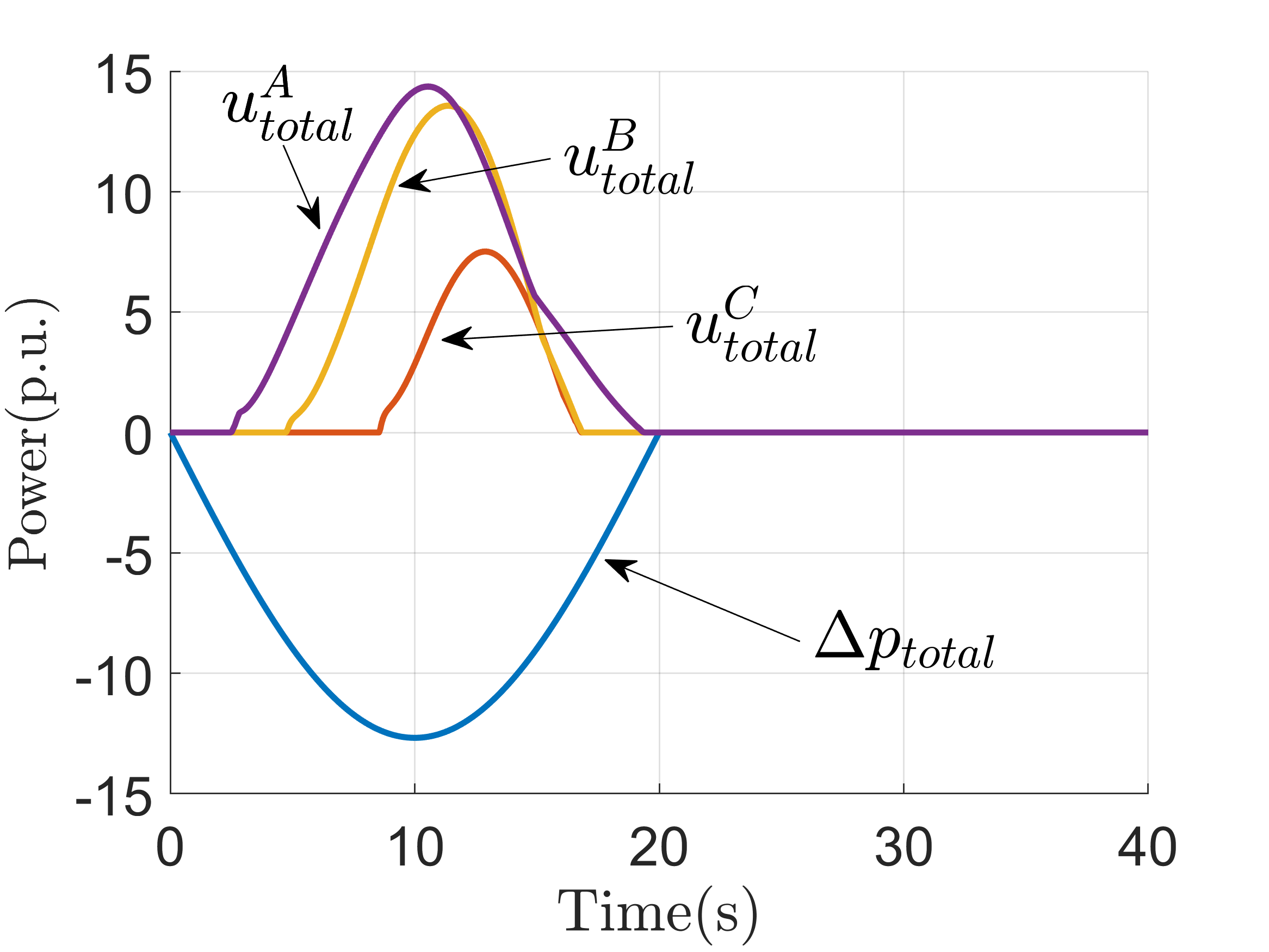}
  \caption{Control signal trajectories with different transient
    frequency bounds. As the frequency bounds become tighter, the
    control signal behaves more alike the negative of power injection
    deviation for more accurate
    compensation.}\label{fig:power-vs-control}
\end{figure}

To illustrate the dependence of the control signal on the tightness of
transient frequency bounds, we perform a simulation where we replace
the frequency bound $\bar\omega_{i}=-\underline\omega_{i}=0.2$Hz by
$\bar\omega_{i}=-\underline\omega_{i}=0.1$Hz and
$\bar\omega_{i}=-\underline\omega_{i}=0.05$Hz for every
$i\in\II{\omega}$. Also, we choose
$\bar\omega_{i}^{\text{thr}}=-\underline\omega_{i}^{\text{thr}}=\bar\omega_{i}/2$
in each case.  Figure~\ref{fig:power-vs-control} shows the overall
power injection deviation $\Delta
p_{total}\triangleq\sum_{i\in\II{}}(p_{i}-p_{i}(0))$ and overall
control signal $u_{total}\triangleq\sum_{i\in\II{u}}u_{i}$ for the
above three cases, where for clarity, we add superscripts $A,B,C$
corresponding to $0.05$Hz, $0.1$Hz, and $0.2$Hz, respectively.  Note
that the control signal trajectory is larger with tighter frequency
bounds, and its shape mimics the trajectory of the power injection
deviation to compensate for it. The overall control signal are
$\int_{0}^{40}u_{total}^{A}\text{d}t=123.2$,
$\int_{0}^{40}u_{total}^{B}\text{d}t=90.1$, and
$\int_{0}^{40}u_{total}^{C}\text{d}t=36.5$ whereas the power deviation
is $\int_{0}^{40}\Delta p_{total}=-161.5$, suggesting significantly
less required control effort as the frequency bound becomes looser.

\begin{figure}[tbh!]
  \centering
  \subfigure[\label{frequency-response-with-control-generator-region1-delayed}]{\includegraphics[width=.32\linewidth]{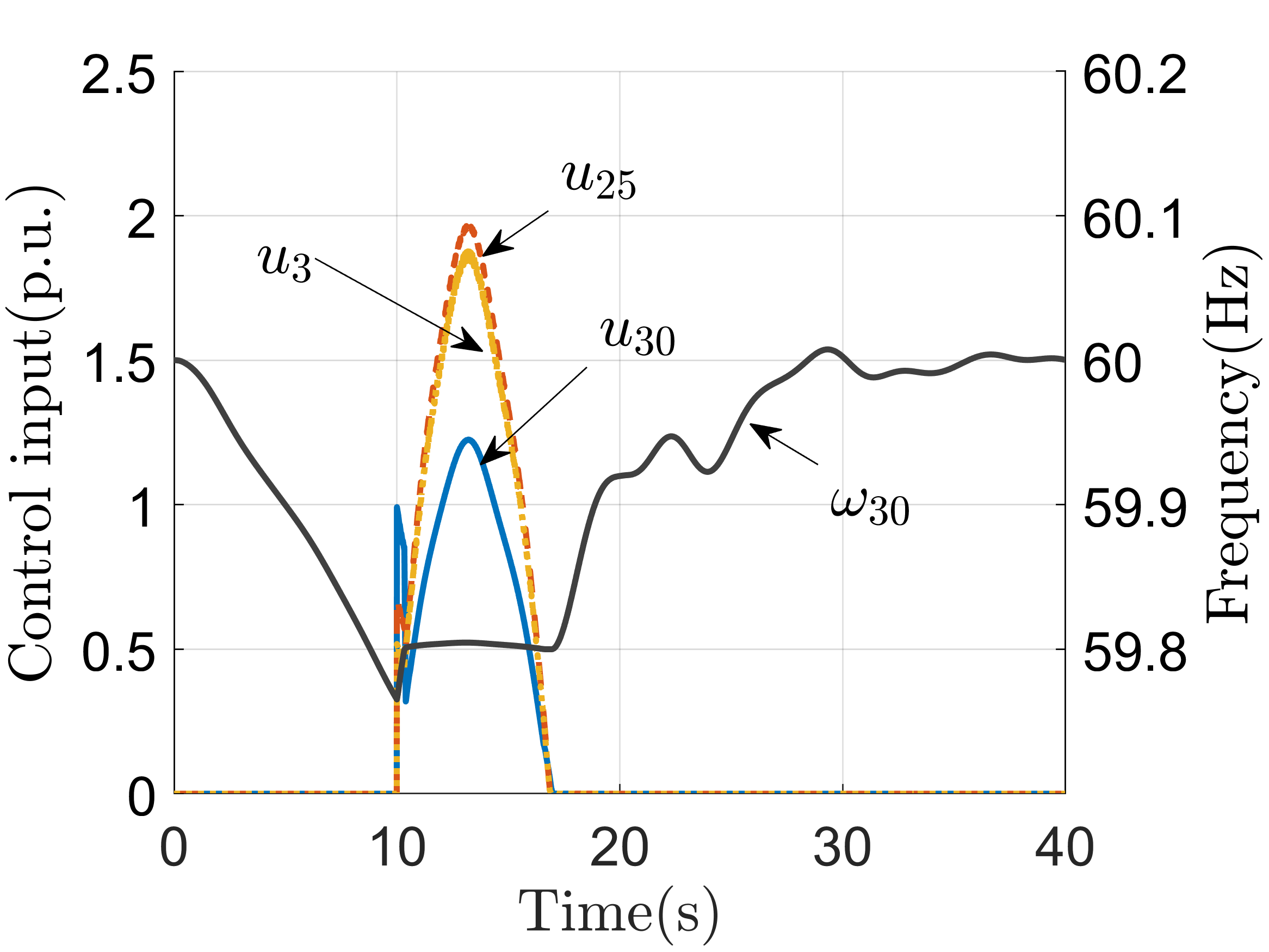}}
  \subfigure[\label{frequency-response-with-control-generator-region2-delayed}]{\includegraphics[width=.32\linewidth]{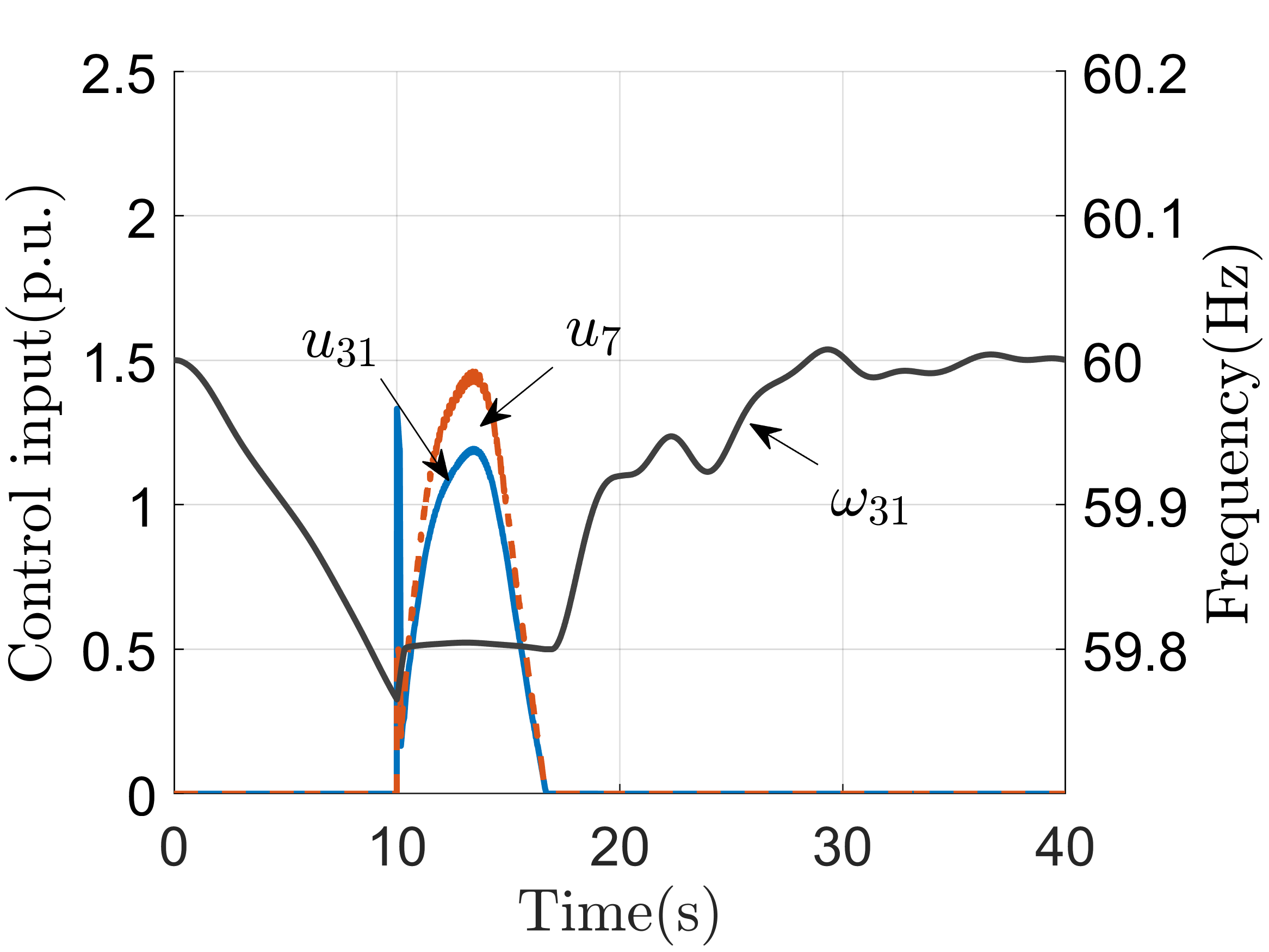}}
    \caption{Frequency and control input trajectories with centralized
      controller available only after $t=10s$,  plot (a) for the region with
      generator 30, and plot (b) for the region with generator~31.
    }\label{fig:trajectories-delayed}
\end{figure}

Next, we simulate the case where generator frequencies are initially
outside the safe frequency region to show how the controller brings
the frequencies back to the safe region.  We apply the same setup used
in Fig.~\ref{fig:trajectories}, but only enable the controller after
$t=10s$. The plots in Fig. \ref{fig:trajectories-delayed} shows the
frequency trajectories and control trajectories of each region. Note
that both two frequency trajectories are lower than 59.8Hz at $t =
10s$. However, as the controller becomes active after $t=10s$, they
come back to the safe region and never leave, in accordance with
Proposition~\ref{prop:dis-control}\ref{item:frequency-attractive-dis}.

\begin{figure}[htb]
  \centering%
  \includegraphics[width=.65\linewidth]{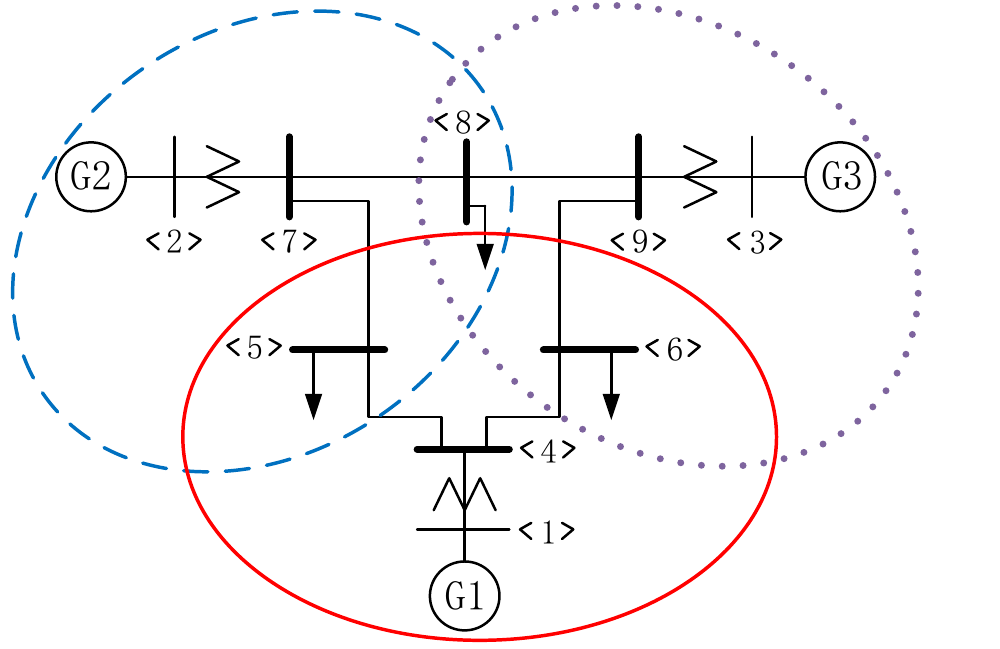}
  \caption{IEEE 9-bus power network with network partition.}\label{fig:IEEE9bus}
\end{figure}

Next, we compare the performance of the centralized
controller~\eqref{eqn:control-input-mpc}, the distributed
controller~\eqref{eqn:control-input-mpc-dis}, and the controller we
proposed in~\citep{YZ-JC:19-auto} in the IEEE 9-bus network with the
regional partition shown in Fig.~\ref{fig:IEEE9bus}.  Since the
control framework in~\citep{YZ-JC:19-auto} requires that controllers
are available only for nodes with transient frequency constraints, for
fairness, we let $\II{\omega}=\II{u}=\{1,2,3\}$ for
controllers~\eqref{eqn:control-input-mpc}
and~\eqref{eqn:control-input-mpc-dis} (adding nodes with controllers
to $\II{u}\slash\II{\omega}$ would further enhance their
performance). We employ a similar set-up as in the previous
simulation, here with $T=0.01s$; $p_{i}(t)\equiv p_{i}(0)$ for
$i=1,2,3$, and $p_{i}(t)=(1+\delta(t))p_{i}(0)$ for $i=4,5,\cdots,9$,
with the coefficient $0.25$ replaced by $1.5$ in $\delta(t)$ so that
the open-loop frequency responses exceed the safe frequency bounds.


\begin{figure}[tbh]
  \centering
  \subfigure[\label{input-centralized}]{\includegraphics[width=.32\linewidth]{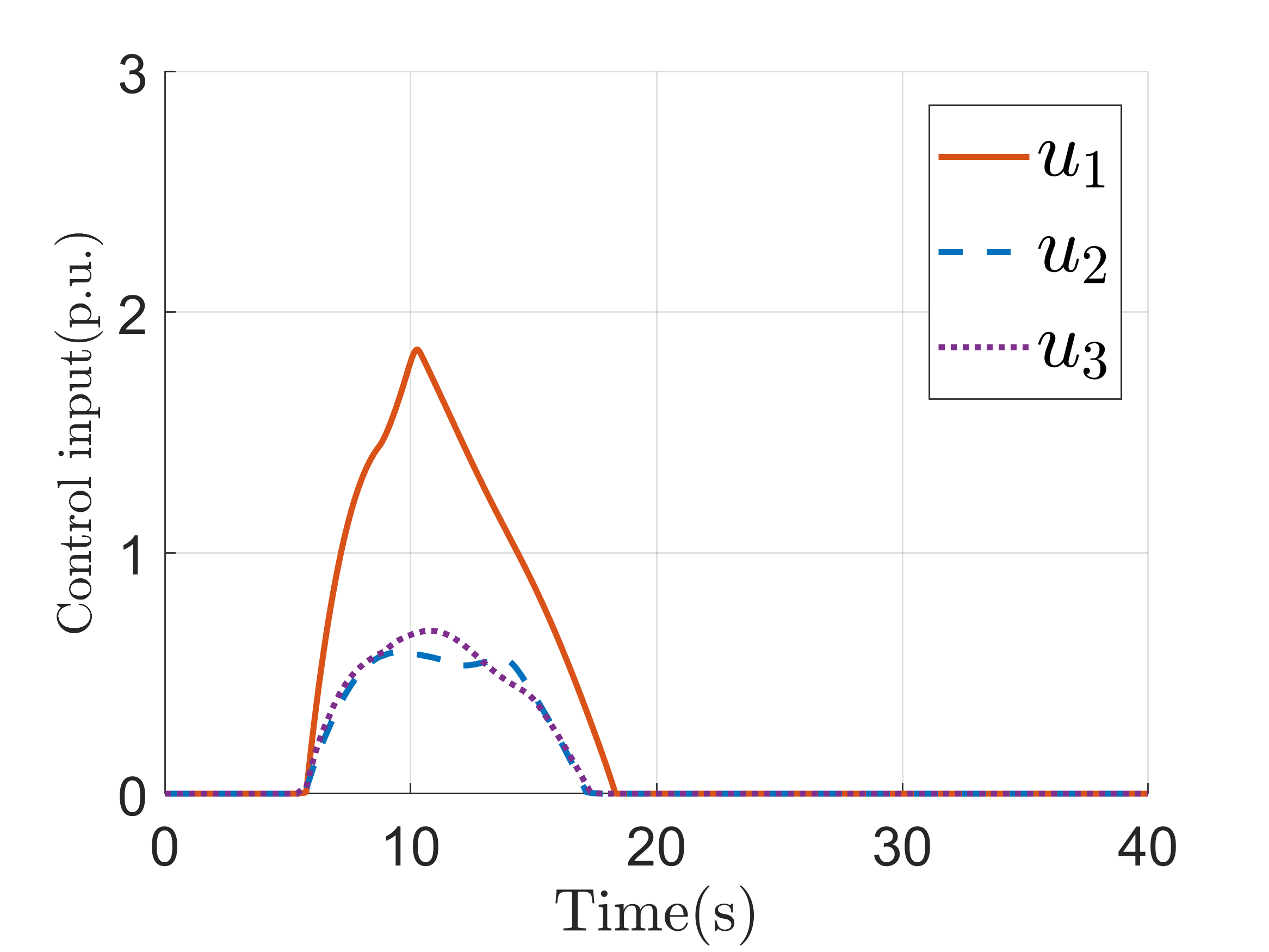}}
  \subfigure[\label{input-distributed}]{\includegraphics[width=.32\linewidth]{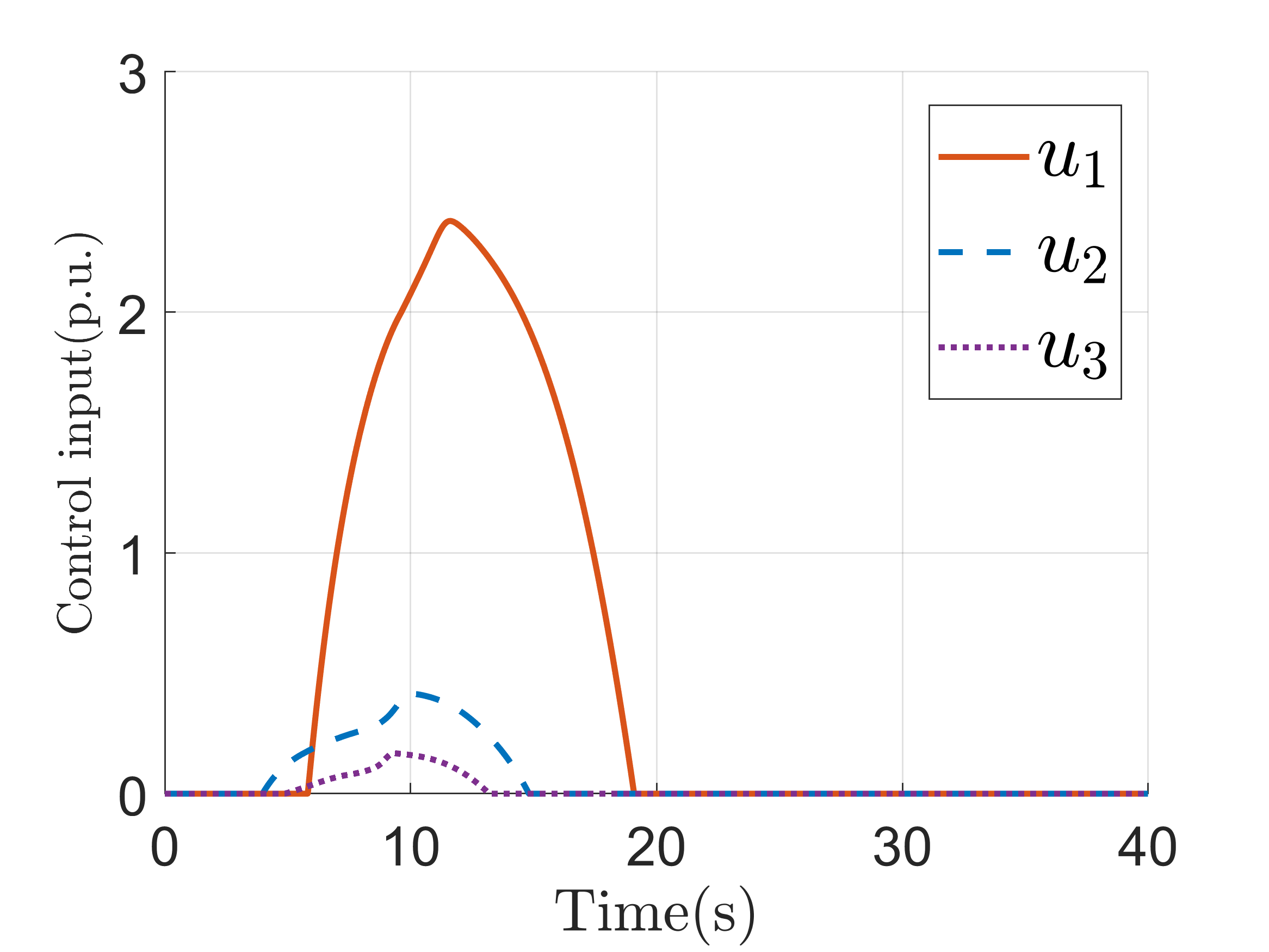}}
  \subfigure[\label{input-df}]{\includegraphics[width=.32\linewidth]{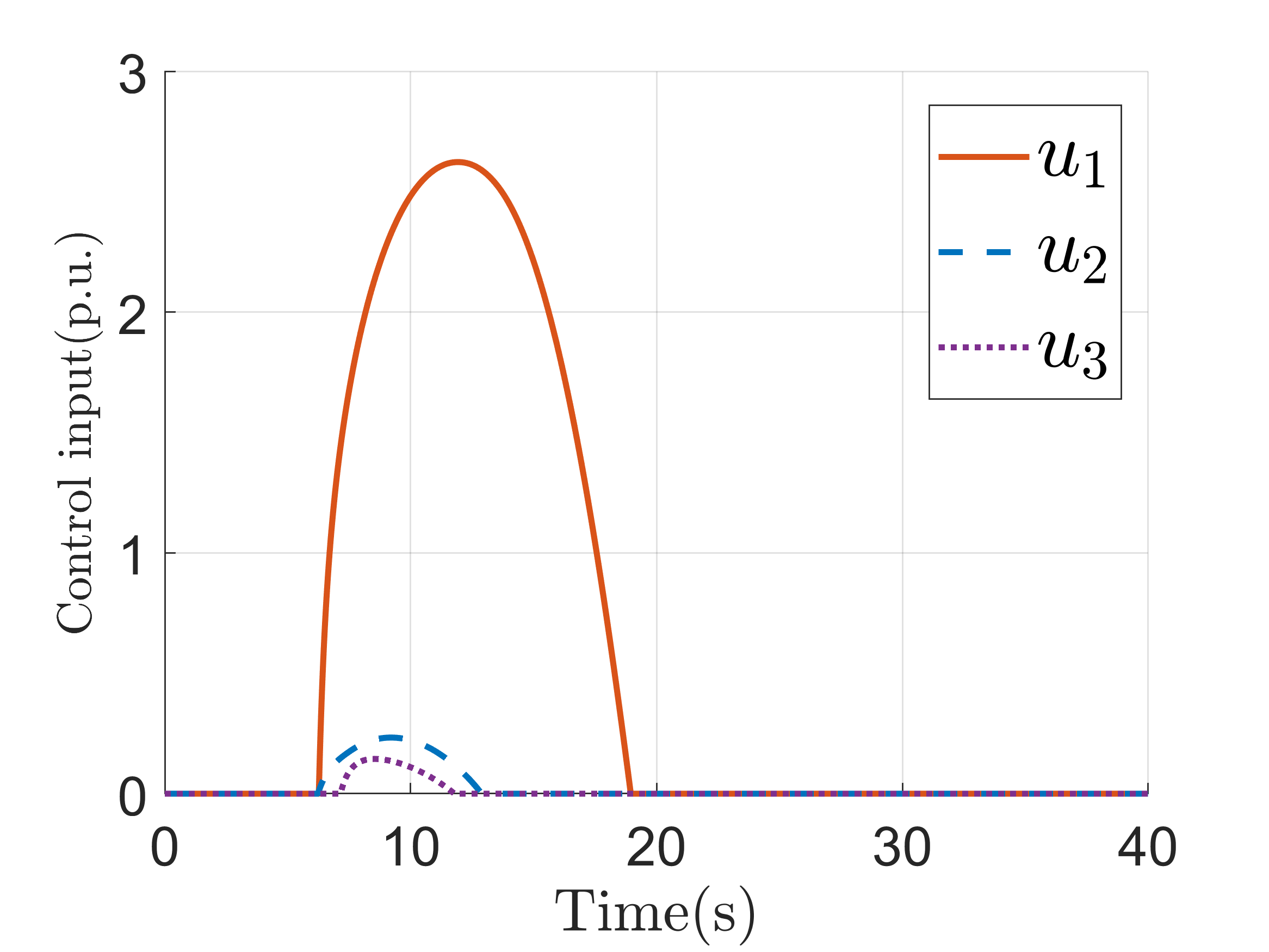}}
  \caption{Input trajectories of controlled generators in IEEE 9-bus
    example under     (a) centralized controller, (b) distributed
    controller, and (c) controller
    proposed in~\citep{YZ-JC:19-auto}. All of them guarantee stability
    and frequency invariance.
  }\label{fig:IEEE-9-trajectories}
\end{figure}

Fig.~\ref{fig:IEEE-9-trajectories} shows the input trajectories of the
generators indexed from 1 to 3 for each of the three
controllers. Since all of them achieve frequency invariance and
stabilization, we do not show the state trajectories. In terms of the
overall control cost, the centralized controller performs the best,
due to its capability of accessing the entire network parameters,
state, and power injection information, and hence all three generators
cooperatively reduce the total cost. This capability is, however,
weakened in the distributed controller, as the controller in each
region only considers its regional optimality, losing inter-region
cooperation.  The controller from~\citep{YZ-JC:19-auto}, which is not
designed by optimizing control effort, tends to have the largest
cost. On the hand, in terms of implementation, the centralized
controller requires global network information as well as solving a
large-scale optimization problem. In comparison, the distributed
controller only accesses network information within its region, and
solves a small-scale optimization problem. The controller
in~\citep{YZ-JC:19-auto} can be computed the fastest and only needs
information of  1-hop neighbors.

\section{Conclusions}
We have proposed centralized and distributed model predictive
controllers for nonlinear power networks that ensure stability and
safe frequency invariance.
We have shown that the closed-loop system preserves the equilibrium
point and local convergence properties of the open-loop system, and
that the control input vanishes in finite time.  
Future work will quantify the loss in optimality incurred by the
convexification of the open-loop optimization problem and the
distributed control framework, study the trade-offs between
discretization accuracy, reference trajectory qualification, and
computational complexity, and analyze the effect of network properties
on the performance and characteristics of the proposed controllers.

{\small
\bibliographystyle{plainnat}%
\bibliography{alias,JC,Main,Main-add}
}
        
\end{document}